\documentclass{emulateapj}
\usepackage{color}
\usepackage{graphicx}
\usepackage{dcolumn}
\usepackage{bm}
\begin{document}
\shorttitle{Partially-depleted galaxy cores}
\title{Sizing up partially-depleted galaxy cores}

\shortauthors{Dullo \& Graham}
\author{Bililign T.\ Dullo\altaffilmark{1},  Alister W.\ Graham\altaffilmark{1}}
\affil{\altaffilmark{1} Centre for Astrophysics and Supercomputing, Swinburne University
of Technology, Hawthorn, Victoria 3122, Australia; Bdullo@astro.swin.edu.au}

\begin{abstract}
We have modelled the inner surface brightness profiles of 39
alleged `core' galaxies with the core-S\'ersic model, and provide new
physical parameters for the largest ever sample of `core' galaxies fit
with this model. When present, additional nuclear components were
simultaneously modelled and the typical rms scatter of the fits (out
to $\sim$10$\arcsec$) is 0.02 mag arcsec$^{-2}$.
Model-independent estimates of each core's break radius are shown to
agree with those from the core-S\'ersic model, and a
comparison with the Nuker model is provided. We found an absence of cores in
what amounts to 18\% of the sample which are reclassified here as
S\'ersic galaxies with low values of $n~(\la 4$) and thus shallow
inner profile slopes. In general, galaxies with $n<3$ and $\sigma <
183$ km s$^{-1}$ do not have depleted cores. We derive updated
relations between core-S\'ersic break radii, their associated surface
brightness, bulge luminosity, central velocity dispersion, and
predicted black hole mass for galaxies with depleted cores. With the
possible exception of NGC 584, we confirm that the inner negative
logarithmic profile slopes $\gamma$ are $\la$ 0.3 for the `core'
galaxies, and $0 > \gamma > -0.1$ for six of these. Finally, the
central stellar mass deficits are found to have values typically
within a factor of 4 of the expected central black hole mass.

\end{abstract}

\keywords{
  galaxies: elliptical and lenticular, cD ---  
 galaxies: fundamental parameter --- 
 galaxies: nuclei --- 
galaxies: photometry---
galaxies:structure  
}

\section{Introduction}\label{Sec_Intro}

The stellar distributions in galaxies have played a valuable role in
guiding our understanding of the galaxies themselves.  In particular,
the accessibility of high-resolution imaging offered by the
\emph{Hubble Space Telescope} (\emph{HST}) substantially advanced our
appreciation of the complexity of galaxy cores (e.g.\ Crane et
al.\ 1993; Kormendy et al.\ 1994; Jaffe et al.\ 1994; Ferrarese et
al.\ 1994; Grillmair et al.\ 1994; van den Bosch et al.\ 1994; Lauer
et al.\ 1995; Byun et al.\ 1996; Gebhardt et al.\ 1996; Carollo et
al.\ 1997; Faber et al.\ 1997). For instance, the centers of real
galaxies may contain such distinct components as bright active
galactic nuclei (AGN), nuclear star clusters, flattened nuclear discs
and bars, dust lanes and clouds.  On the other hand, giant stellar
evacuation zones are also observed.  Luminous galaxies with such
shallow cores had of course long been known to exist from ground-based
observations (e.g.\ King \& Minkowski 1966, 1972; King 1978; Young et
al.\ 1978; Binney \& Mamon 1982; see the review by Graham 2012a) but
\emph{HST} enabled us to accurately quantify these.

After studying 14 bright elliptical galaxies with the pre-refurbished
\emph{HST}/WFPC1, Ferrarese et al.\ (1994) introduced a 4-parameter double power law model
% (with no transition controlling parameter) 
to describe the inner surface brightness distributions of bright
galaxies. While the (relatively brighter) galaxies in their sample
which possessed depleted cores with shallow inner profiles were
grouped as ``Type I", the remaining galaxies, labeled ``Type II", had
a profile that remained steep all the way into the center.  Examining
a larger sample of galaxies imaged using the same \emph{HST}/WFPC1
high-resolution Planetary Camera, Kormendy et al.\ (1994) and Lauer et
al.\ (1995) largely agreed with the division of
galaxies presented in Ferrarese et al.\ (1994), referring to them as `core' galaxies and `power-law'
galaxies, respectively.  They also advanced a double power law model
which they dubbed the `Nuker law' for fitting the (underlying host
galaxy) surface brightness profiles of early-type galaxies. The Nuker
model had an additional fifth parameter to moderate the transition
between the two power laws --- as introduced by Hernquist (1990, his
eq.\ 43) for modelling the internal density profiles of galaxies.

The physical process(es) responsible for the observed difference
between the inner surface brightness profiles of `core' galaxies and
the fainter `power-law' galaxies (nowadays referred to as `S\'ersic'
galaxies because these spheroids have S\'ersic light profiles rather
than power-law light profiles) provide valuable clues about the
galaxies' past history. In bright galaxies, the widely advocated `dry'
(i.e.\ gas poor) galaxy merger hypothesis (e.g. Faber et al.\ 1997)
can result in the gravitational sling shot of central stars (core
scouring) due to the coalescence of supermassive black holes (SMBHs)
from the progenitor galaxies (e.g.\ Begelman, Blandford \& Rees 1980;
Makino \& Ebisuzaki 1996; Merritt \& Milosavljevi\'c 2005; Merritt
2006). It is possible that the sizes and mass deficits of such partially depleted cores may
 reflect the amount of merging and damage caused by the
black holes (after having eroded any pre-existing nuclear stellar
components: Bekki \& Graham 2010). Having an accurate quantification
of the physical parameters defining the centers of galaxies is
therefore important. Moreover, reliable break radii $R_b$, used to
denote the sizes of the cores, may even be useful for predicting black
hole masses (Lauer et al.\ 2007a).

While investigating the lack of any connection between the double
power-law model and the curved galaxy brightness profiles observed
outside of the cores, Graham et al.\ (2003, see their figures~2--4)
revealed that the Nuker model's break radius, and other parameters,
were not robust quantities but are sensitive to the radial range of
the surface brightness profile that is fitted. For example, the break
radii $R_b$ were shown to vary by more than a factor of three. The
parameters' sensitivity was recognised to arise from the Nuker model's
efforts to fit an outer power-law to what is actually a curved
brightness profile. The luminosity profiles of bright (core) galaxies
($M_{B} \la -20.5 $ mag) --- which show a downward deviation from the
inward extrapolation of their outer S\'ersic (1963, 1968) profile ---
were subsequently shown to be precisely represented by the
core-S\'ersic model (Graham et al.\ 2003; Trujillo et al. 2004).

While Lauer et al.\ (2005) missed this development, Ferrarese et
al.\ (2006) found the core-S\'ersic model to be highly applicable to
bright early-type galaxies in the Virgo cluster.  Lauer et
al.\ (2007a,b) subsequently wrote that ``Graham et al.\ (2003) have
criticized the Nuker $r_b$ as being sensitive to the domain over which
the Nuker law was fitted, particularly when the outer limit of the fit
extends only slightly beyond $r_b$. In practice, however, the Nuker
laws are fitted over a large radial range that extends well beyond
$r_b$.''
However this was not the problem identified by Graham et 
al.\ (2003), who had demonstrated that the Nuker model parameters
deviated further from the true values as the fitted radial extent was
{\it increased}. 
 
Based on ``work in preparation'' Lauer et al.\ (2007b) refuted that
their Nuker break radii were biased ``in any way'' because their radii
reportedly agreed very well with model-independent values of where the
curvature in the surface brightness profile was a maximum.  This was a
surprising claim because these latter values should not be dependent
on the radial extent of the data while the Nuker model break radii
{\it are} a strong function of the fitted radial extent (Graham et
al.\ 2003).
Kormendy et al.\ (2009, their section 4.1) subsequently buoyed the
Nuker model and dismissed the core-S\'ersic model.
G\"ultekin et al.\ (2009) then over-looked any and all concerns about
the Nuker model which they presented along with Nuker model parameters
from Lauer et al.\ (2005), and encouraged readers to use this data,
additionally noting that the surface brightness profiles that were fit
with the Nuker model are available at the Nuker web
page\footnote{http://www.noao.edu/noao/staff/lauer/wfpc2\_profs/}.
G\"ultekin et al.\ (2011) continued in this vein, motivating us to
further investigate, nearly a decade on, the issue of whether the
Nuker model parameters are reliable, physically meaningful quantities,
or if instead the core-S\'ersic model parameters may be preferable.
At stake is not only the accuracy to which we quantify the cores of
galaxies, but our subsequent understanding of cores and how they
relate to their galaxy at large.

In this paper, we focus on the nuclear structure of galaxies by
re-analyzing the surface brightness profiles of all 39 `core' galaxies
imaged with the WFC2 / F555W or F606W filter and listed in Lauer et
al.\ (2005) to be a `core' galaxy (see section~\ref{Sec_Data}).
For reference, Trujillo et al.\ (2004) modelled only 9 possible `core'
galaxies, Ferrarese et al.\ (2006) modelled 10, and Richings, Uttley
\& Kr$\ddot{\textrm{o}}$ding (2011) have very recently modelled 21
`core' galaxies. We are therefore modelling the largest sample of
suspected `core' galaxies to date.
For comparison's sake with the Nuker model break radii, we use exactly
the same surface brightness profiles as Lauer et al.\ (2005),
available at the previously mentioned Nuker web-page.

We first concentrate on measuring the core size using the core-S\'ersic model
(see sections~\ref{Sec_cS} and \ref{Sec_Fit}). We additionally take the
nuclear excess, usually nuclear star clusters or AGN emission, into account
while modeling the underlying host galaxy light. In section~\ref{Sec_Rb} we
use two model-independent core size estimators and reveal that one of these
can not be used while the other is consistent with our core-S\'ersic break
radii.
Furthermore, we confirm
that the published Nuker model break radii are typically 100\% larger than the
break in the surface brightness profile determined relative to the
inward extrapolation of the outer S\'ersic function (Trujillo et al.\ 2004).
We additionally report that `artificial' break radii have been reported in
what were alleged to be `core' galaxies but are actually S\'ersic galaxies
with no break in their S\'ersic profile and which thus have no partially
depleted core relative to their outer light profile (section~\ref{Sec_core}).
Throughout this paper we use terms such as 'actual', `true' and `real' break
radii and cores when referring to galaxies that have inner surface brightness
profiles which break downward from (i.e.\ have lower flux than) the inward
extrapolation of the outer S\'ersic model which describes their outer stellar
distribution.

Sets of structural parameter relations encompassing central as well as
global properties are presented in section~\ref{Sec_rel}. In
particular, equations involving the break radius and associated
surface brightness, and the luminosity, are derived. We also
investigate the (core size)-(central black hole mass) relation in
section~\ref{Sec_Rb-BH}. Using updated data, we find that the break radius {\it can} be 
used to consistently predict the black hole mass when using either the
$M_{bh}$-$\sigma$ or $M_{bh}$-$L$ relations for `core' galaxies
when coupled with our updated $R_{b}$-$\sigma$ and $R_{b}$-$L$
relations. We go on to discuss the detection of additional nuclear
components in the full sample of S\'ersic and core-S\'ersic galaxies
in section~\ref{Sec_Add} while section~\ref{Sec_Con} summarises our
main conclusions.

\section{Sample Selection}\label{Sec_Data}

Lauer et al.\ (2005) analyzed and presented fits to the major-axis
surface brightness profiles of 77 relatively bright, nearby,
early-type galaxies.  Every galaxy in their sample was observed with
the Wide Field Planetary Camera 2 (WFPC2: Biretta et al.\ 2001)
onboard the \emph{HST} and was centered on the PC CCD (which has an
image scale of 0\arcsec.0456 pixel$^{-1}$ and a field view of 800
$\times$ 800 pixels). Their sample lacks any characterizing selection criteria
and comprises galaxies at distances of $\sim $ 10--100 Mpc. While
almost all of the `core' galaxies were imaged with the F555W filter
(similar to broadband \emph{V}), two were not and we use their F606W
data (roughly broadband \emph{R}) instead. We refer the reader to Lauer et al.\ (2005) for an extensive description of
the sample and images, which includes procedures adopted for PSF
deconvolution\footnote{Some of the merits and disadvantages of image
deconvolution are described in Ferrarese et al.\ (2006).}, dust obscuration
correction, sky subtraction and background source masking.

\begin{center}
\begin{figure*}
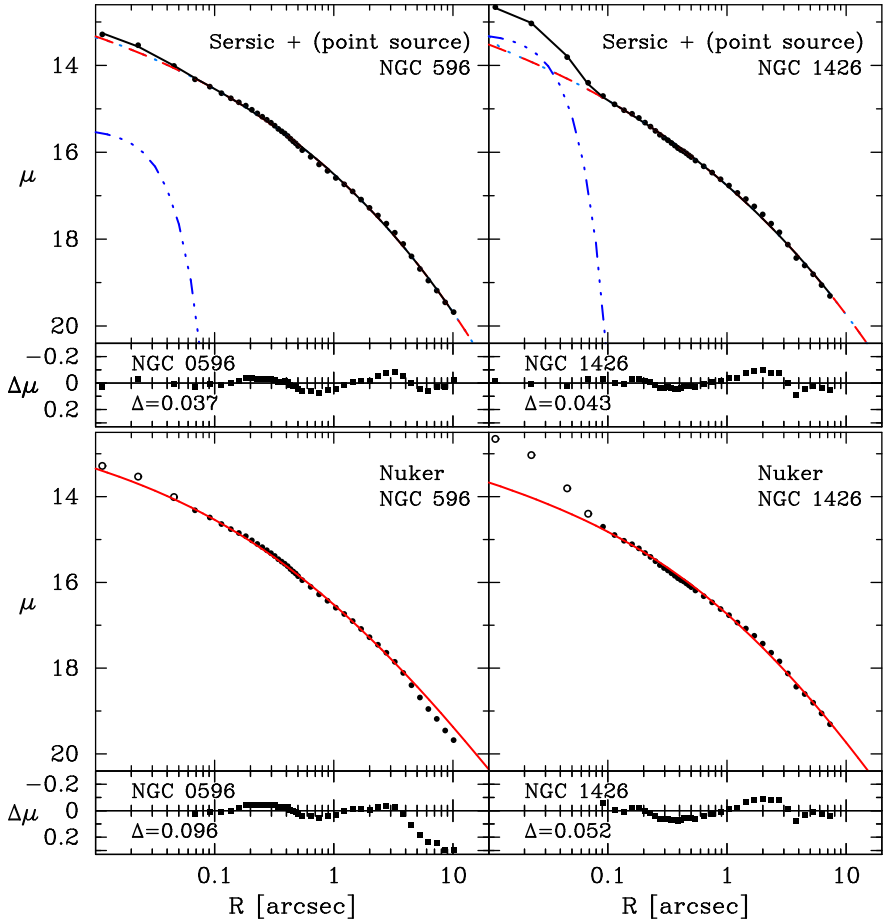

\includegraphics[angle=270,scale=.70]{csboth.ps}\\
\includegraphics[angle=270,scale=.70]{nukerboth.ps}

\caption{Left panel: 
A 3-parameter S\'ersic plus 2-parameter point-source model describes the
entire available radial extent of NGC 596 while the 5-parameter Nuker model
(fit taken from Lauer et al.\ 2005) fails to describe the inner and outer
light profile. 
Right panel: A 3-parameter S\'ersic plus 2-parameter point-source model
describes NGC~1426 better than the 5-parameter Nuker fit taken from
Lauer et al.\ (2005) which can not describe the nucleus.  The rms scatter, 
$\Delta$, about the major-axis, F555W-band surface brightness profiles pertains
only to the data points included in the fit (shown by the filled symbols).
}
\label{Fig1}
\end{figure*} 
\end{center}

Our sample comprises 39 early-type galaxies which are the `core' galaxy subset
of the 77 galaxies presented in Lauer et al.\ (2005).  This sample selection enables a direct
comparison with the (published) Nuker model's estimation of the core size and
related parameters, and to further achieve this direct comparison we have
modelled the same published light profiles$^1$. The global properties
of our target galaxies are
 summarized in Table 1, which
presents their morphology, magnitude, updated distance and velocity
dispersion. 
For the six lenticular galaxies plus one Sa spiral galaxy, we have roughly 
converted their total galaxy magnitudes, reported by Lauer et al.\ (2007b), into
bulge magnitudes using a mean $V$-band bulge-to-disc ratio of 1/3, equivalent to a 
mean bulge-to-total ratio of 1/4 (Graham \& Worley 2008; Laurikainen et al.\
2010, their section~6.3 and references therein).  The 1$\sigma$ range on 
this $B/T$ ratio for S0--Sa galaxies is about a factor of 2, corresponding 
to a 1$\sigma$ uncertainty of $\pm$0.75 mag for our bulge magnitudes. In passing we note that for 
four of these seven disk galaxies (NGC~507, NGC~2300, NGC~3607, and the Sa
galaxy NGC~7213), the 
dust-corrected $K$-band $B/T$ ratios reported by Laurikainen et al.\ (2010) are 0.33,
0.28, 0.33 and 0.18, respectively, suggesting that we are not too far off with our
adopted $B$-band value of 0.25.  Laurikainen et al.\ do however identify NGC 3706 as an elliptical galaxy, 
while de Vaucouleurs et al.\ (1991) refer to it as SA0(rs), 
and we have no comparison $B/T$ ratios for NGC~4382 and NGC~6849. 
In any event, it should be remembered that these latter three 
galaxies make up a fairly small fraction of the total sample. 

\begin{center}
\begin{table}
\begin {minipage}{75mm}
\caption{Updated global parameters of the `core' galaxy sample from Lauer et al.\ (2005).}
\label{Tab1}
\begin{tabular}{@{}llcccccc@{}}
\hline
\hline
 Galaxy&Type& $M_{V}$ & D &$\sigma$ \\
&&(mag)&(Mpc)&(km s$^{-1}$)\\
(1)&(2)&(3)&(4)&(5)\\
\multicolumn{1}{c}{} \\              
\hline                           
NGC 0507        & S0   &  $-21.54$ &$63.7^{n}$ &306\\
NGC 0584         & E     $$  &  $-21.12$ &$19.6^{t}$&206  \\
NGC 0741         & E      $$ &  $-23.31$&  $72.3^{n}$&291\\
NGC 1016         & E    $$  &  $-23.22$ & $88.1^{n}$&302  \\
NGC 1374         & E      $$&  $-20.39$ & $19.2^{t}$&183\\
NGC 1399        & E      $$ &  $-21.89$&  $19.4^{t}$&342\\
NGC 1700          & E      $$  &  $-22.53$ & $53.0 ^{n}$&239\\
NGC 2300         & S0     $$  &   $-19.90$ &  $25.7^{n}$&261\\
NGC 3379         & E      $$  &  $-20.86$ & $10.3^{t}$ &209 \\
NGC 3607         & S0      $$  &  $-19.95$ & $22.2^{t} $  &224\\
NGC 3608         & E      $$  &  $-21.05$ &    $22.3^{t}$&192\\
NGC 3640         & E     $$ &  $-21.80$ &$ 26.3^{t}$    &182\\
NGC 3706        & S0     $$  &  $-20.56$ &$45.2^{n}$ & 270\\
NGC 3842         & E      $$ &  $-23.04$ & $ 91.0^{n} $& 314\\
NGC 4073         & cD     $$  &  $-23.33$ &  $85.3 ^{n}$&275\\
NGC 4278         & E      $$  &  $-20.90$ &    $15.6^{t}$&237\\
NGC 4291         & E      $$ &  $-20.68$ &$ 25.5 ^{t}$ &285\\
NGC 4365        & E     $$ &  $-22.00$ &  $19.9^{t}$  &  256\\
NGC 4382          & S0      $$  &  $-20.47$& $17.9 ^{t}$  &179\\
NGC 4406       & E      $$  &  $-22.31$ &   $16.7^{t}$ &235\\
NGC 4458         & E      $$  &  $-19.13$ & $ 16.8^{t}$  &103\\ 
NGC 4472         & E      $$ &  $-22.66$ &  $15.8^{t} $  &294\\
NGC 4473         & E      $$ &  $-20.82$ &  $ 15.3 ^{t}$ &179\\
NGC 4478        & E     $$ &  $-19.85$ &  $17.6^{t}$  &137\\
NGC 4486B        & cE     $$ &  $-18.77$ &$ 25.8^{n}$  & 170\\ 
NGC 4552         & E     $$ &  $-21.25$ & $14.9^{t}$&253\\
NGC 4589         & E      $$  &  $-21.01$ &$ 21.4^{t}$ &224\\
NGC 4649         & E      $$ &  $-22.32$ &$16.4^{t}$&335\\
NGC 5061         & E      $$  &  $-22.43$ & $32.6^{n}$   &186\\
NGC 5419         & E      $$  &  $-23.27$ &  $59.9^{n}$&351\\
NGC 5557          & E      $$  &  $-22.34$ &$46.4^{n}$&253\\
NGC 5576      & E      &  $-21.11$ & $24.8^{t}$& 171\\
NGC 5813         & E      $$ &  $-22.20$ &$ 31.3^{t}$ &237\\
NGC 5982         & E      $$  &  $-22.04$ &$41.8^{n}$ &239\\
NGC 6849         & SB0      $$  &  $-21.21$ & $80.5^{n}$& 209\\ 
NGC 6876         & E      $$&  $-23.45$ & $54.3^{n}$ & 229 \\
NGC 7213        & Sa     $$  &  $-20.38$ & $21.1^{n}$ &163\\
NGC 7619         & E      $$ &  $-22.76$ & $51.5^{t}$  &323\\
NGC 7785         & E      $$  &  $-21.96$ & $47.2 ^{n}$&255\\
NGC 0596$^{*}$         & cD     $$  &  $-20.81$ &$21.2^{t}$&151\\ 
NGC 1426$^{*}$         & E     $$  &  $-20.51$ &$23.4^{t}$&151 \\
\hline
\end{tabular} 

Notes.---Col. (1) Galaxy name. Col. (2) Morphological classification from the NASA/IPAC Extragalactic Database (NED)\footnote{(http://nedwww.ipac.caltech.edu)}. Col. (3) Absolute \emph{V}-band (galaxy or bulge) magnitude obtained from Lauer et al.\ (2007b) and adjusted using the distance from col. (4). Sources: ($t$) Tonry et al.\ (2001) after reducing their distance moduli by 0.06 mag (Blakeslee et al.\ 2002); ($n$) from NED (3K CMB). Col. (5) Central velocity dispersion from HyperLeda\footnote{(http://leda.univ-lyon1.fr)} (Paturel et al.\ 2003). The superscript * is used to indicate two `power-law' galaxies taken from Lauer et al.\ (2005) and used for illustrative purpose in Fig.~\ref{Fig1}.

\end {minipage}
\end{table}
\end{center}

\section{The S\'ersic and core-S\'ersic models}\label{Sec_cS}

By combining CCD images of early-type Virgo cluster galaxies with deep, large
field-of-view, photographic images, Caon et al.\ (1993) revealed that the
S\'ersic (1963) model fits the main parts of the profiles of both elliptical
and spheroidal galaxies astonishingly well over large ranges in surface
brightness (down to $\sim$29 $B$-mag arcsec$^{-2}$). In general, after excluding the nuclear region, the S\'ersic R$^{1/n}$ model
fits the surface brightness profile of both elliptical galaxies and the bulges
of disk galaxies remarkably well over their entire radial range (Caon,
Capaccioli \& D'Onofrio 1993; D'Onofrio, Capaccioli \& Caon 1994; Young \&
Currie 1994; Andedakis et al.\ 1995; Graham et al.\ 1996). In fact, systematic
deviations from this model appear to signal either the presence of a central
light deficit or an additional nuclear component (Balcells et al.\ 2003;
Graham \& Guzm\'an 2003). Indeed, using CCD images with the extended profiles from Caon et al.\ (1993)
and other photographic data, one of main conclusions noted in Kormendy et al.\ (2009) was exactly this.  

The radial intensity distribution of the 3-parameter S\'ersic $R^{1/n}$ model, a
generalization of de Vaucoulers (1948) 2-parameter $R^{1/4}$ model, is defined as
\begin{equation}
I(R) = I_{e} \exp \left \{ -b_{n} \left[  
\left(\frac{R}{R_{e}}\right)^{1/n}-1\right]\right \},
\end{equation}
where $ I_{e}$ denotes the intensity at the half-light radius $R_{e}$. The
quantity $b_{n}\approx 2n- \frac{1}{3}$, for $1\la n\la 10$ (e.g.\ Caon et
al.\ 1993), is a function of the shape parameter $n$,
and is defined in a way to ensure that $R_{e}$ encloses half of the total
luminosity.  A review of the S\'ersic model and its associated expressions can
be found in Graham \& Driver (2005). 

Many, but not all, spheroids fainter than -20.5 $B$-mag contain additional nuclear components. This is illustrated in Fig.~\ref{Fig1} where we display a Nuker model and a S\'ersic plus point source
model fit to the surface brightness profiles of NGC 596 and NGC 1426.  The
profiles have been taken from the Nuker web-pages, and the fitted Nuker model
parameters are from Lauer et al.\ (2005). Both
models agree on the absence of a partially depleted core, and as such the
galaxies are classified as S\'ersic galaxies. There are, however, considerable
differences regarding the quality of the fits.
For NGC 596, the 3-parameter S\'ersic plus 2-parameter Gaussian model
accommodates the entire observed radial range of the brightness profile
remarkably well with a smaller root mean square (rms) residual than the 5-parameter Nuker
model. The Nuker model fit (taken from Lauer et al.\ 2005) to this galaxy's
light profile not only excluded the inner most data points but also clearly
reveals a significant departure from the profile in the outer region.
Similarly, while the S\'ersic model plus Gaussian function can represent the
entire observed profile of NGC 1426, the 5-parameter Nuker model cannot describe the extra compact light source at the center and has
more scatter in the residual profile.

As mentioned previously, 
the light profiles of luminous (M$_{B} \la -20.5 $ mag) elliptical galaxies depart
systematically from the S\'ersic model near their center. It is important to
realize that this departure, a downward deviation with respect to the inward
extrapolation of the outer S\'ersic profile, emanates from a central starlight
deficit and is not due to dust (which would result in a dramatic color
change).  Such stellar distributions can be described using the core-S\'ersic
model introduced by Graham et al.\ (2003) and applied in Trujillo et al.\
(2004). A blend of an inner power-law and an outer
S\'ersic function, it can be written as 
\begin{equation}
I(R) =I' \left[1+\left(\frac{R_{b}}{R}\right)^{\alpha}\right]^{\gamma /\alpha}
\exp \left[-b\left(\frac{R^{\alpha}+R^{\alpha}_{b}}{R_{e}^{\alpha}}
\right)^{1/(\alpha n)}\right], 
 \end{equation}
with 
\begin{equation}
I^{\prime} = I_{b}2^{-\gamma /\alpha} \exp 
\left[b (2^{1/\alpha } R_{b}/R_{e})^{1/n}\right].
\end{equation}
$I_{b}$ is intensity at the core's break radius $ R_{b}$, $\gamma$ is the
slope of the inner power law region, and $\alpha$ controls the sharpness of the
transition between the inner power-law and the outer S\'ersic profile.  As in
the S\'ersic model, $R_{e}$
is the effective half-light radius of the outer S\'ersic function and $b$ has the
same general definition as before\footnote{ One recovers the S\'ersic
R$^{1/n}$ function from equation (2) when setting R$_{b}$ and $\gamma $ to
zero.}. In practice, the 6-parameter core-S\'ersic model can be reduced to a
5-parameter model by setting $\alpha $ to some large constant value.  Trujillo
et al.\ (2004) set $\alpha\rightarrow \infty$, so that the transition from
S\'ersic profile to power law at $R_{b} $ is infinitely sharp, with no
transition region, an approach effectively adopted by Ferrarese et al.\ (2006).
In this paper, as in Richings et al.\ (2011), 
we explore and use a range of finite values for $\alpha$.

\section{Fitting Analysis}\label{Sec_Fit}

We fit the one dimensional light distributions of the 39 underlying host
galaxies using S\'ersic and core-S\'ersic models, and account for any
additional nuclear light components with either a Gaussian or an exponential
model. Fits to the full 39 galaxies are available in Appendix~A. The quality of the fits, as indicated by the rms given in each panel of Appendix~A Fig.\ 21, is excellent, typically 0.01-0.03 mag arcsec$^{-2}$. 

As can be seen, the 3-parameter S\'ersic model proffers a good match to two (NGC 4473 and NGC
5576) of the 39 galaxies\footnote{The Nuker model parametrization (Lauer et
al.\ 2005) reported that NGC 4473 and NGC 5576 had break radii of 4.45 and
4.18 arcseconds, respectively.} all the way to the \emph{HST} resolution
limit.
While these galaxies have shallow central profiles, the profiles do not
`break' from the outer envelope --- further evidenced by the small values of $\alpha$
used with the Nuker model (Lauer et al.\ 2005).
Application of the S\'ersic model, along with small inner Gaussian and
exponential functions, yields a satisfactory fit to the luminosity profiles of an additional
five nucleated galaxies (NGC 1374, NGC 4458, NGC 4478, NGC 4486B and NGC
7213).
In what follows, and as noted before, we collectively refer to this class of galaxy without
depleted cores as `S\'ersic' galaxies, as done by Trujillo et al.\ (2004) who
first reported that NGC~4458 and NGC~4478 are S\'ersic galaxies, i.e.\ they do
not display any \emph{downward} departure from the S\'ersic $ R^{1/n}$ profile
at their centres. Kormendy et al.\ (2009) and Hopkins et al.\ (2009) also identify central light excesses over the S\'ersic function in these two galaxies. 
Except for NGC~4486B (Lauer et al.\ 1996) and NGC~7213, the rms residual
scatter is $\la$ 0.032 mag arcsec$^{-2}$ for these seven S\'ersic galaxies.
These two exceptions are two of only four galaxies, from the full sample of
39, with complicated structure (see Appendix~A Fig.~\ref{Fig18}). As shown in section 6, these seven galaxies stand out from the `core'-galaxies in a number of systematic ways.

\begin{figure}
\includegraphics[angle=270,scale=.60]{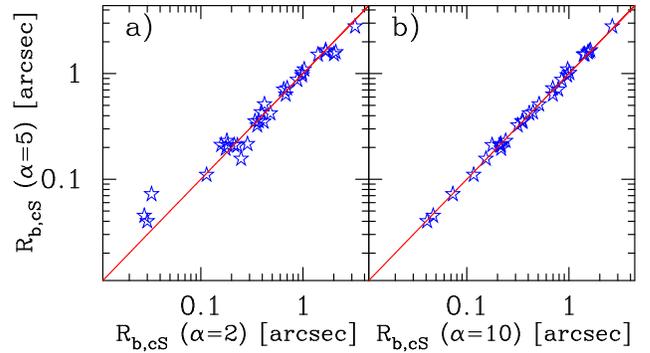}
\caption{Robustness of the core-S\'ersic break radii. Break radii
from moderate-transition, $\alpha=5$, core-S\'ersic fits plotted against a) break
radii from broad-transition fits, $\alpha=2$ and b) sharp-transition fits, $\alpha=10$. The Pearson correlation coefficient ($r$) between $R_{b,cS}$ ($\alpha=5$) and $R_{b,cS}$ ($\alpha=2$) is 0.99, while $r$ of  $R_{b,cS}$ ($\alpha=5$) versus  $R_{b,cS}$ ($\alpha=10$) is 1.0. }
\label{Fig2}
\end{figure}

There is some variation in $\alpha$ from galaxy to galaxy when using the
core-S\'ersic model. The role of the parameter $\alpha$ is to moderate the
sharpness of the transition between the outer S\'ersic profile and the inner
power-law, with higher values corresponding to sharper transitions and vice
versa. The profile of NGC~4291, for example, has a
sharper transition and hence requires a larger value of $\alpha$ than say 
NGC~1399.  We have set $\alpha = 10, 5$ and 2 for matching sharp,
moderate and broad transition regions respectively. The change in the core
size, i.e.\ the break radius, while varying $\alpha$ has been closely
inspected. 
The robustness of the break radius is illustrated by the
plots shown in Fig.~\ref{Fig2}, which compare the break radii obtained using different
values of $\alpha$. The Pearson correlation coefficient ($r$) between the core-S\'ersic break radius $R_{b,cS}(\alpha=10)$ and the core-S\'ersic break radius $R_{b,cS}(\alpha=5)$ shown in Fig.~\ref{Fig2}b is 1.0. Although the break
radii from the broad-transition model (i.e.\ using $\alpha=2$) has a small
amount of scatter in Fig.~\ref{Fig2}a, this is due to the fact that the
majority of our sample galaxies do not prefer the broad-transition model.
There are, however, a small handful of galaxies (NGC 1016, NGC 1399, NGC 2300,
NGC 3379, NGC 4365, NGC 4649, NGC 5419 and NGC 5813) whose profiles have a
broad transition region (see Table~\ref{Tab2}).

We also estimate the uncertainties of the core-S\'ersic model parameters by exploring their stability for different $\alpha$ values (i.e. 2, 5 and 10). In agreement with Richings et al.\ (2011) and Trujillo et al.\ (2004), for some galaxies, we notice that $\gamma$, $n$ and $R_{e}$ obtained from the best $\alpha=5$ and 10 fits are slightly different from the ones which are obtained using $\alpha=2$.

\begin{figure}
\includegraphics[angle=270,scale=.520]{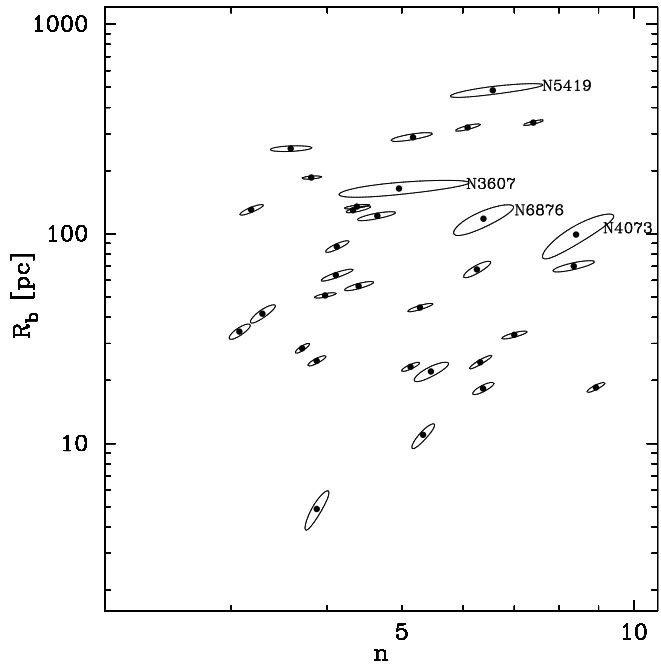}
\caption{Parameter coupling between the core-S\'ersic break radius $R_{b}$ and S\'ersic index $n$ for our `core' galaxy sample. The contours show the $\Delta \chi^{2}=1$ boundaries, and their projections onto the horizontal and vertical axes give the 68.3\% (1$\sigma$) confidence intervals around each galaxies' optimal $n$ and $R_{b}$ values, respectively. Each contour is generated by holding all the core-S\'ersic parameters, other than $R_{b}$ and $n$, fixed at their best-fit values (Table 2). }
\label{Figchi_sq}
\end{figure}

In addition, we explore the coupling between the core-S\'ersic break radius $R_b$ and the S\'ersic index $n$ by holding the other parameters constant at their best-fit values (Table 2) and using the $\chi^{2}$ distribution, which we normalised at the minimum value. In Fig.~\ref{Figchi_sq} we show the 68.3\% (1$\sigma$) confidence limits around the optimal $R_b$ and $n$ values for the core galaxies using the $\Delta\chi^{2}=1.0$ contours after marginalizing over the remaining core-S\'ersic parameters. In general, Fig.~\ref{Figchi_sq} indicates the absence of a coupling between $R_b$ and $n$, as seen by the small contours. Overall, we estimate the uncertainties associated with $\gamma$, $R_{b}$, $n$ and $R_{e}$ to be roughly 10\%, 10\%, 15\% and 20\%  respectively. We also note that the errors in the S\'ersic parameters could partly be associated with the limited radial extent of our data (see section 4.1). 

As noted in passing above, 
only 4 galaxies are somewhat poorly represented using our models 
(NGC 4073, NGC 4486B, NGC 6876 and NGC 7213: see Appendix~A, Fig.~\ref{Fig18}). 
NGC~7213 is a Seyfert (Sa) galaxy with a winding nuclear dust spiral which can
be traced all the way to the nucleus (Deo, Crenshaw \& Kraemer 2006). It
appears that this prominent dusty nuclear feature in the galaxy profile (see
Fig.~\ref{Fig3}), which the core-S\'ersic model is not designed to recover,
is the likely origin for the residual pattern about the model fit to this
galaxy, particularly in the inner R $\la$ 0$\arcsec$.3 region.
The cD galaxy, NGC 4073, has an inner ring (Lauer et al.\ 2005) over the
$0\arcsec.1< R < 0.\arcsec4$ region and we exclude these few data points from
the fit.  
NGC~6876 is a dominant
elliptical galaxy in the Pavo group with a possible past or ongoing merger
history (Machacek et al.\ 2005).  The residual structures outside the core
regions of NGC~4073 and NGC~6876 that are seen in Appendix A's Fig.~\ref{Fig18} seem to
be associated with the change in ellipticities of these galaxies as presented
by Lauer et al.\ (2005). 
Finally, although not apparent from the \emph{I} and \emph{V} broadband
\emph{HST}/WFPC2 image, Lauer et al.\ (1996) remarked on the presence of a
double optical nucleus from the deconvolved WFPC2 image of NGC 4486B. This
creates a spurious depleted core as noted by Lauer et al.\ (1996), see also
Soria et al.\ (2006) and Ferrarese et al.\ (2006).

\begin{figure*}
\includegraphics[angle=0,scale=0.458]{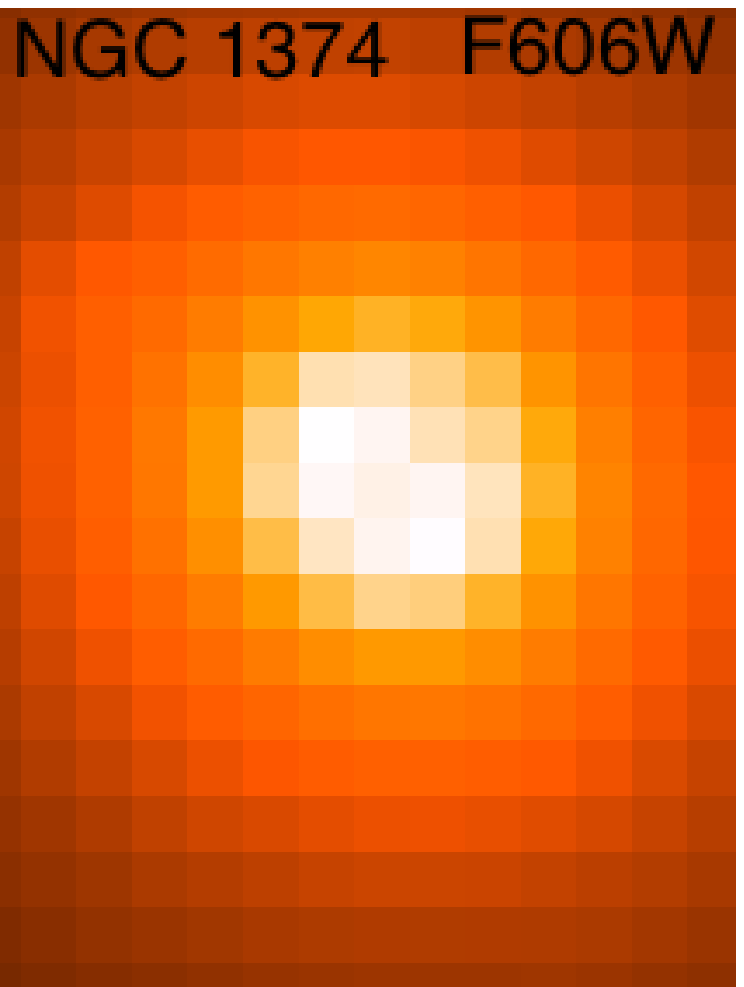}
\includegraphics[angle=0,scale=0.31]{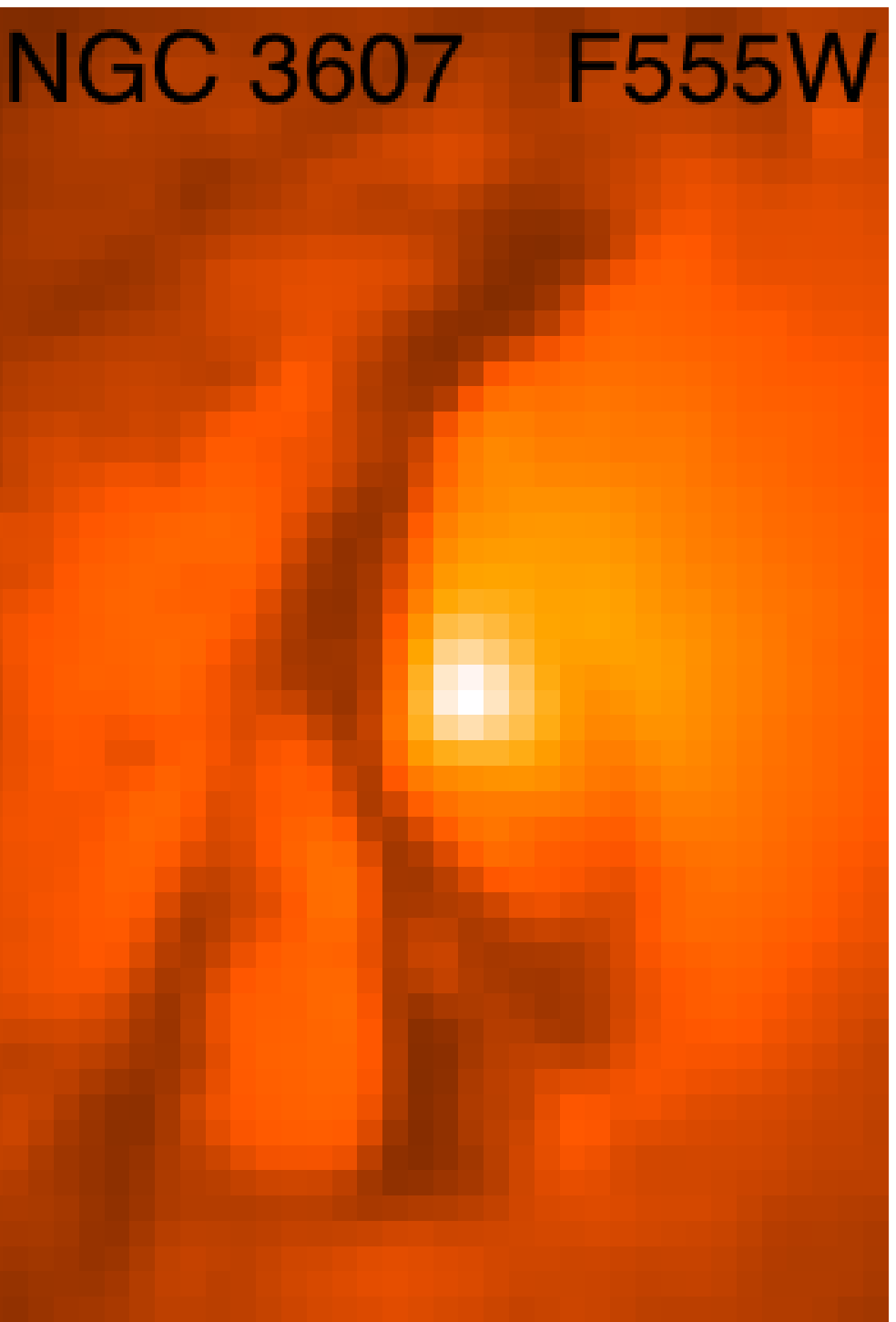}
\includegraphics[angle=0,scale=0.445]{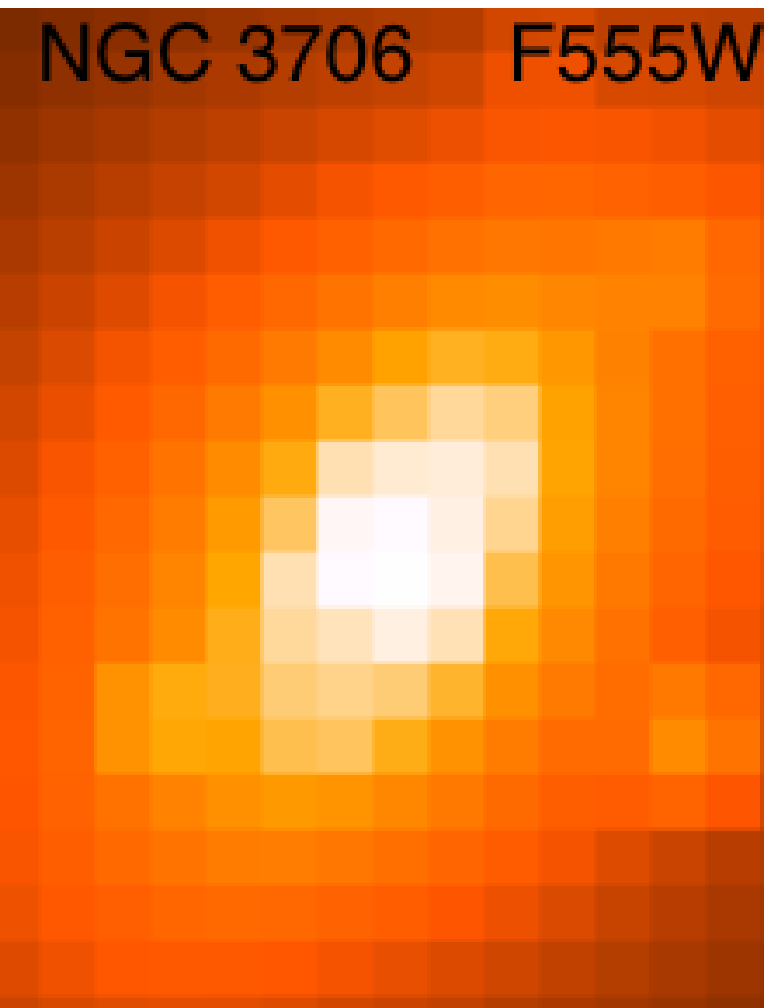}
\includegraphics[angle=0,scale=.395]{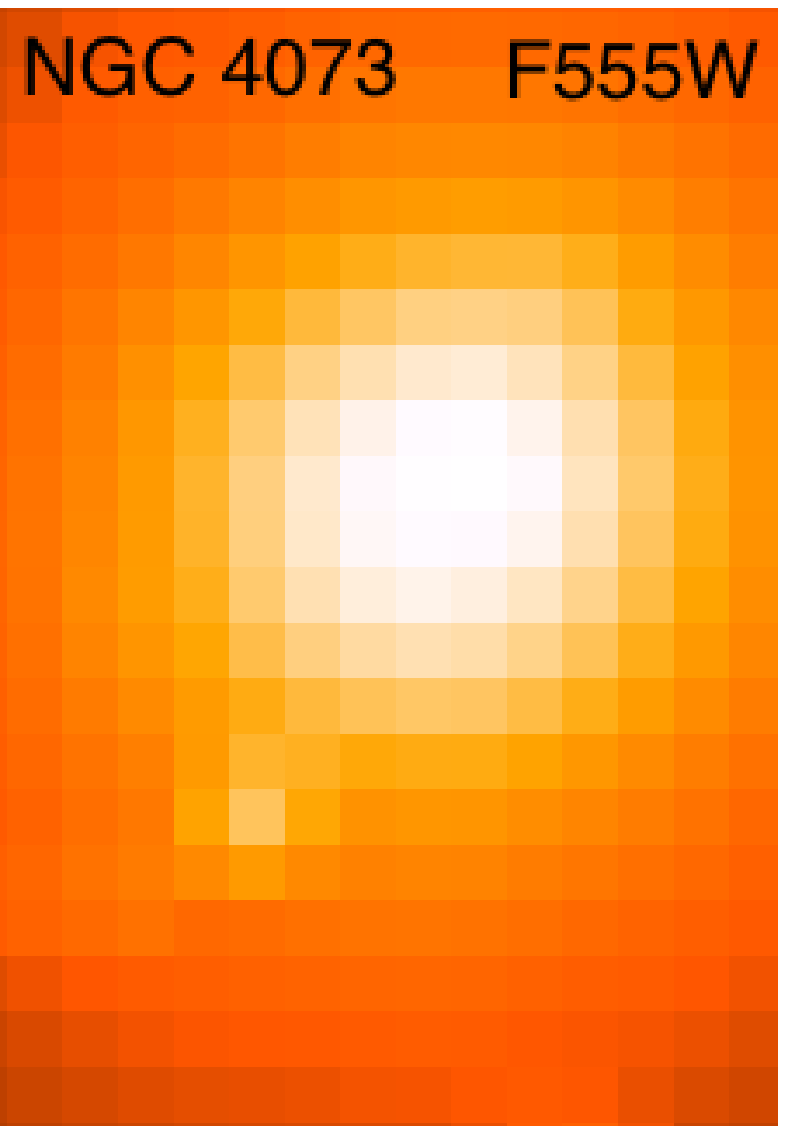}
\includegraphics[angle=0,scale=0.39]{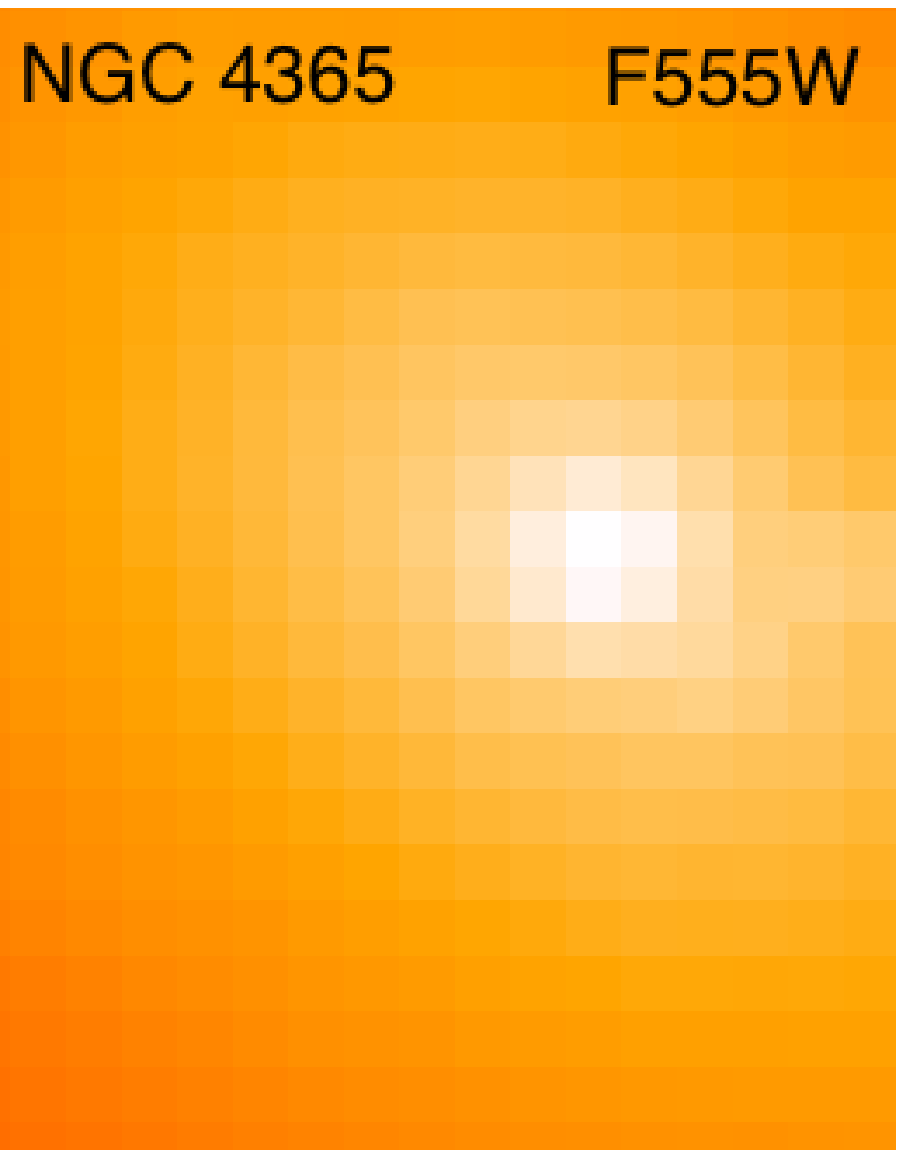}
\caption{Optical \emph{HST }images of galaxy centres. Each galaxy was observed
using the high-resolution (0$\arcsec$.046/pixel) PC1/WFPC2 camera.}
\label{Fig3}
\end{figure*}
\setcounter{figure}{3}
\begin{center}
\begin{figure*}
\includegraphics[angle=0,scale=0.43]{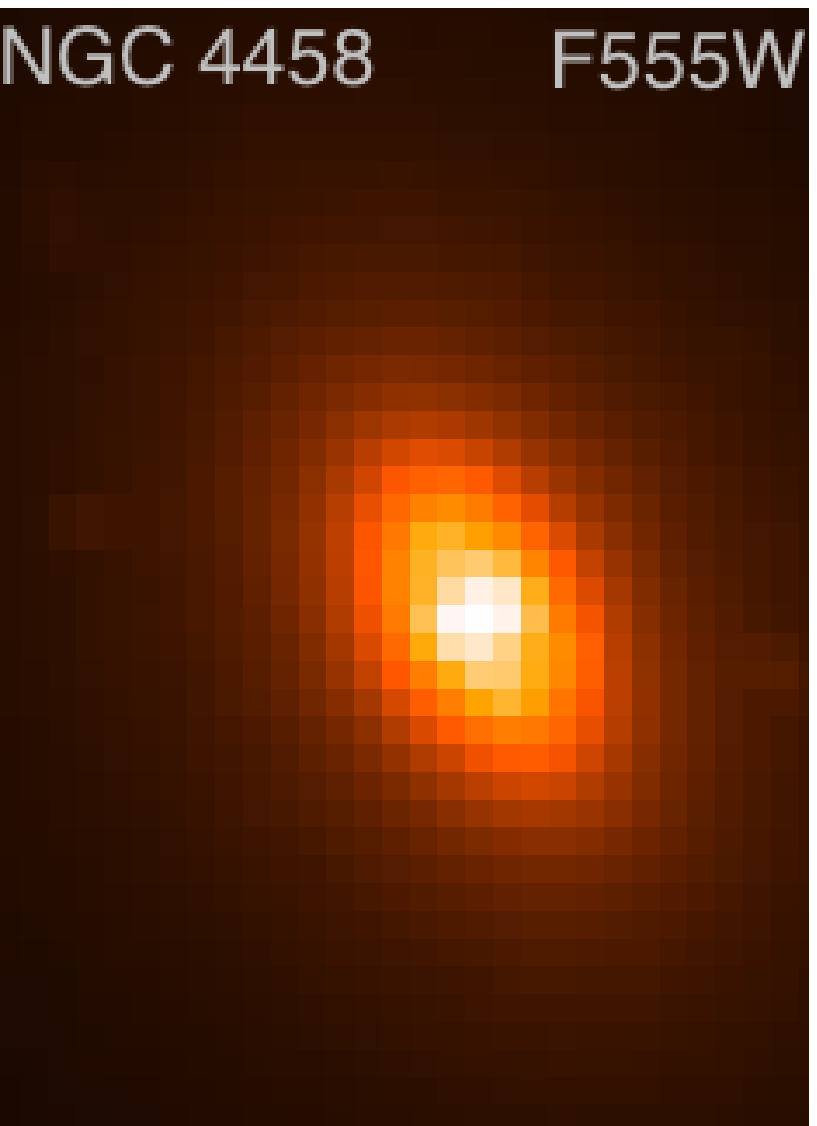}
\includegraphics[angle=0,scale=.43]{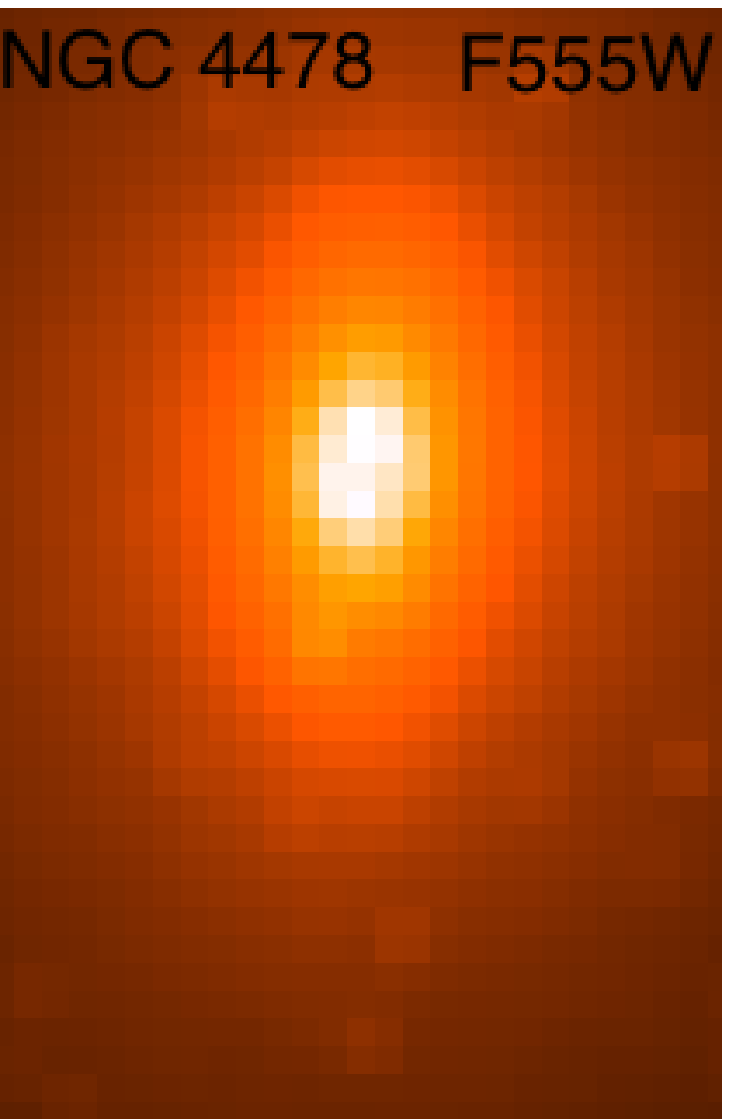}
\includegraphics[angle=0,scale=.44]{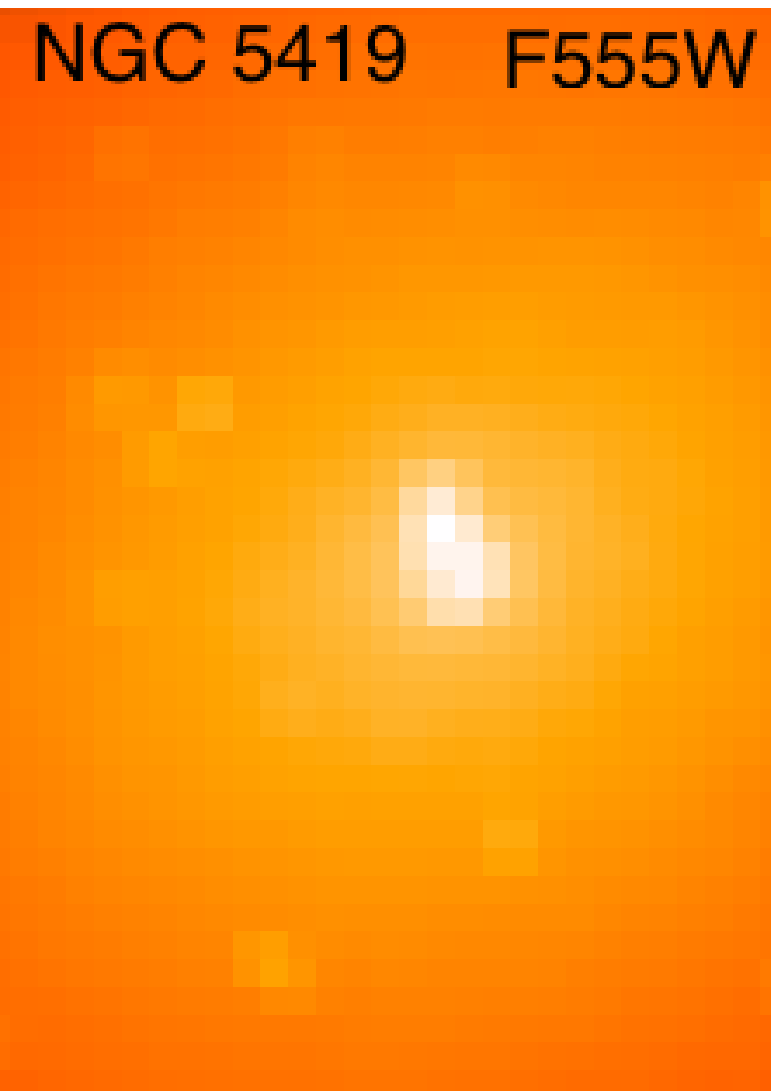}
\includegraphics[angle=0,scale=.43]{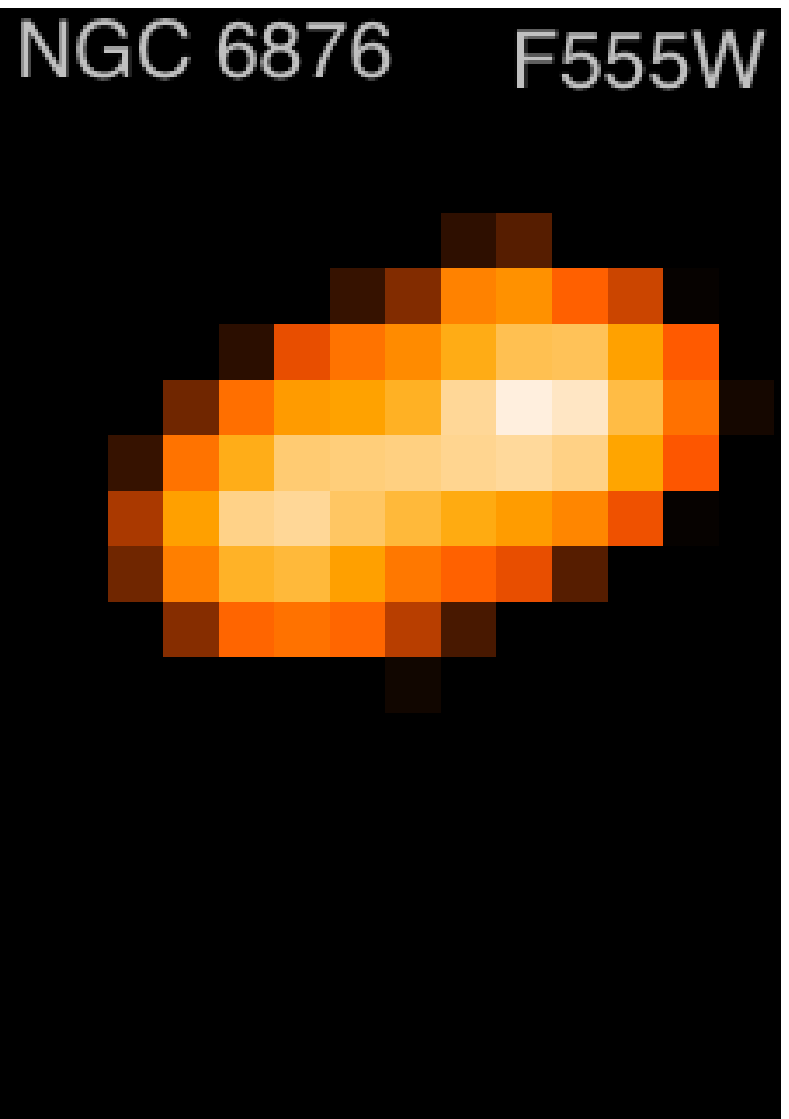}
\includegraphics[angle=0,scale=.425]{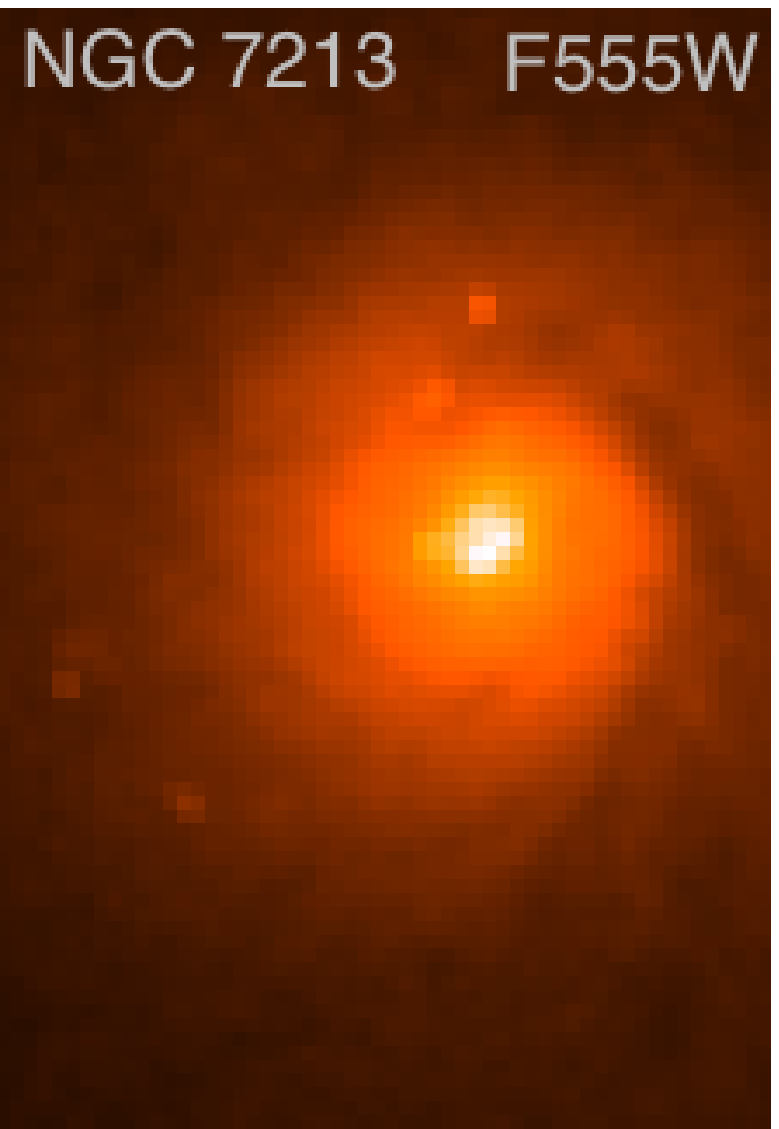}
\caption{continued}
\end{figure*}
\end{center}

\begin{table*}
\begin {minipage}{185mm}
~~~~~~~~~~~\caption{Structural parameters}
\label{Tab2}
\begin{tabular}{@{}llccccccccccc@{}}
\hline
\hline
Galaxy&$ \mu_{b, V} $ & $R _{b}$ &$R_{b}$ &$ \gamma$&$\alpha$&$n$&Profile Type&$m_{pt, V}$&$\mu_{0, V}$&$h$&Notes\\
&mag arcsec$^{-2}$&(arcsec)&(pc)&&&&&(mag)&mag arcsec$^{-2}$&(arcsec)\\
(1)&(2)&(3)&(4)&(5)&(6)&(7)&(8)&(9)&(10)&(11)&(12)\\
\multicolumn{6}{c}{} \\ 
\hline                                           
NGC 0507   &16.45     &  0.42  &  130&  $ 0.09$ &5&3.2&$c$-$S$    \\
NGC 0584    & 14.57 & 0.19 & 18 & 0.46     &5 &6.4&$c$-$S$&&     \\
NGC 0741    &17.52     &  0.96   &338&  $ 0.19$    &5 &7.4& $c$-$S$&22.4&&  & & \\
NGC 1016    &17.21     &   0.68 &289  &  $ 0.20$  &  2 & 5.2& $c$-$S$  \\
NGC 1374$^{+}$  &--- &   --- &  ---&  ---  & --- &2.8& $S$&&14.3&0.17& \\
NGC 1399$^{+}$   &16.29      &     2.09&  $196$ &  $ 0.11$     &2 &4.0& $c$-$S$ \\
NGC 1700    & 13.64     &    0.07&  18  &  $ 0.27$  &5   & 8.9&   $c$-$S$     \\
NGC 2300     &16.85 &  0.98&122&   $ 0.11$  &   2& 4.7& $c$-$S$  \\
NGC 3379     &15.66    &   1.03  &  51&  $ 0.19$    & 2& 4.0&     $c$-$S$    \\
NGC 3607     &16.42&1.52 & 164& 0.29&5&5.0& $c$-$S$ &&&&nuclear dust lanes    \\
NGC 3608   &15.10     &     0.21&  23 & 0.28& 5&5.1&$c$-$S$  \\
NGC 3640   &14.72      &     0.04 &5 & -0.02 & 5&3.9&  $c$-$S$  \\
NGC 3706   &14.17     &    0.11& 24 & -0.02 &10&6.3&$c$-$S$  &&&&ring of stars (0\arcsec.06-0\arcsec.4) \\
NGC 3842  &17.42      &    0.72& 320  & 0.19 & 5& 6.1&  $c$-$S$    \\
NGC 4073     &16.47    &   0.24& 99  & -0.05    & 10&8.4&  $c$-$S$  &&&&ring of stars (0\arcsec.1-0\arcsec.4)   \\
NGC 4278     &15.79    &    0.75 & 56  & 0.20     &5& 4.4&    $c$-$S$&19.4 &&& \\
NGC 4291      &15.22   &     0.35&44 & 0.10  & 5 &5.3&    $c$-$S$\\
NGC 4365    &16.56&    1.40& 135& 0.04    &2&4.4& $c$-$S$&20.1&&& \\
NGC 4382    &15.04      &  0.32 & 28 & 0.08   & 5& 3.7&   $c$-$S$     \\
NGC 4406    &16.01  &   0.87&70 & 0.01     &5& 8.4&   $c$-$S$   \\
NGC 4458      &---&  ---&  ---  &  ---     &---&3.1&   $S$  &17.0& 14.4&0.25& \\
NGC 4472  &16.44     &  1.68 & 129& 0.00 & 2&4.3 & $c$-$S$&22.2&&&  \\
NGC 4473   &  --- &  ---  &  ---&  ---    &---&2.1&  $S$ \\
NGC 4478   &  --- &    --- & --- &  ---     &---&2.7& $S$&20.1&15.6&0.41& \\
NGC 4486B     & ---  &     ---& --- &  ---    & ---& 3.0&$S$&&& &double optical nuclei   \\
NGC 4552    &15.00    &    0.35 & 25 & 0.01& 10&3.9& $c$-$S$& 20.5&&& \\
NGC 4589       &15.34 &     0.21&   22& 0.30   & 5&5.5&  $c$-$S$    \\
NGC 4649    &16.92   &     3.23&256 &  0.18   &  2&3.6& $c$-$S$    \\
NGC 5061    &14.06    &     0.21& 33&  0.13    &5&7.0 &  $c$-$S$   \\
NGC 5419      &17.60   &    1.67&  485& -0.05   &   2 &6.6&  $c$-$S$&19.9&&& \\
NGC 5557      &15.31    &     0.16&  35 & 0.17 & 5&3.1 &$c$-$S$  \\
NGC 5576   &--- & ---& ---&  --- & ---& 3.5&   $S$   \\
NGC 5813      &  16.15 &     0.42& 64& -0.09 &  2&4.1& $c$-$S$      \\
NGC 5982        &15.45 & 0.21    & 42  & 0.08   & 5  &3.3&   $c$-$S$  \\
NGC 6849         &16.72& 0.22    &  84& 0.20   &5&6.3  & $c$-$S$ \\
NGC 6876         &16.98& 0.45   & 119& -0.01  &   10    & 6.4&  $c$-$S$ &&&&double optical nuclei  \\
NGC 7213        &---& ---   &  ---&  ---    &---&1.5& $S$ &16.6&12.6&0.04& \\
NGC 7619        &15.78 &    0.35 &  87 &  0.12  &5&4.1& $c$-$S$        \\
NGC 7785       &14.94& 0.05 & 11 &0.16  &10&5.3& $c$-$S$&\\
\hline
\end{tabular} 

Notes.---Structural parameters from fits to the $V$-band major-axis surface
brightness profiles (Appendix A Figs.~\ref{Fig17} and \ref{Fig18}). 
The superscript + is used to indicate that an F606W surface brightness profile
is used, rather than an F555W surface brightness profile.
Col. (1) Galaxy
name. Col. (2)-(7) Best-fit parameters from the core-S\'ersic model,
equation (2). Col. (8) Indicates the profile classification where \emph{c-S} =
core galaxy described by the core-S\'ersic model, and \emph{S}= S\'ersic
galaxy described by the S\'ersic model. Col. (9) Point source apparent
magnitude. Col. (10) Nuclear disk central surface brightness. Col. (11) Nuclear
disk scale length. Col. (12) Description of inner additional light
components. A ``?'' indicates a tentative classification. 

\end{minipage}
\end{table*}

\subsection{Literature comparison of core-S\'ersic fits}\label{Sec_Lit}

We have seven galaxies (NGC 1700, 4291, 4458, 4478, 5557, 5576 and 5982) 
in common with Trujillo et al.\ (2004) and 10 galaxies 
(NGC 4365, 4382, 4406, 4458, 4472, 4473, 4478, 4486B, 4552 and 4649) 
in common with Ferrarese et al.\ (2006) and C\^ot\'e et al.\ (2006).  
While Trujillo et al.\ (2004) classified NGC 1700 as a S\'ersic galaxy after
fitting a profile sampled from $R$ $\sim$ 0$\arcsec$.1 to $ 70\arcsec.0$, 
we {\it tentatively} identify a small core within $R_b \sim 0\arcsec.07$. Apart
from NGC 1700, our profile classifications are in agreement with
Trujillo et al.\ (2004) and Ferrarese et al.\ (2006).  

\begin{figure}
\includegraphics[angle=270,scale=.50]{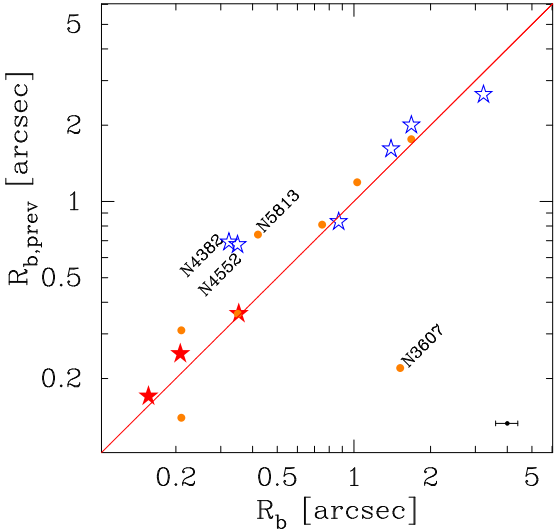}
\caption{
Comparison of the core-S\'ersic model's break radii from this study ($R_b$: 
Table 2, $V$-band) and previous break radii, $R_{b,prev}$, from 
i) Trujillo et al.\ (2004, $R$-band, filled stars), ii) Ferrarese et al.\ (2006, $g$-band, open stars) and iii) Richings et al.\ (2011, various bands, filled circles). 
We have converted the geometric mean radii from Ferrarese et al.\ (2006) into semi-major axis radii using their 
ellipticity values. A Representative error bar is shown
at the bottom of the panel.
}
\label{Fig4_a}
\end{figure}

\begin{figure}
\includegraphics[angle=270,scale=.50]{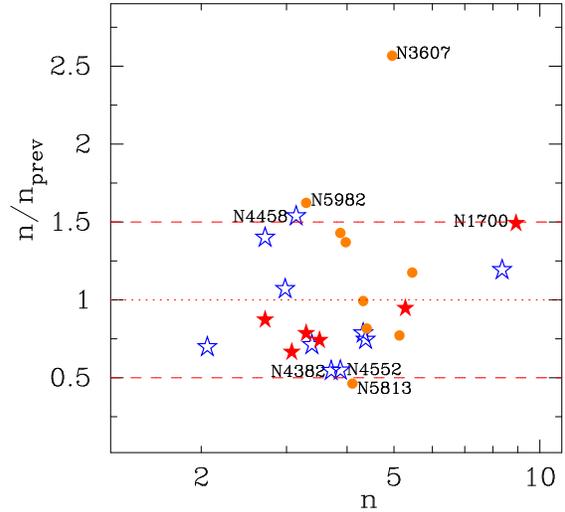}
\caption{
Comparison of our major-axis S\'ersic indices, $n$, with previously published 
values ($n_{prev}$) derived using a greater radial extent.  The filled stars and circles are major-axis
S\'ersic indices from Trujillo et al.\ (2004) and Richings et al.\ (2011) respectively while the open stars are geometric-mean axis S\'ersic indices from Ferrarese et al.\ (2006). 
}
\label{Fig4_b}
\end{figure}

\begin{figure}
\includegraphics[angle=270,scale=.50]{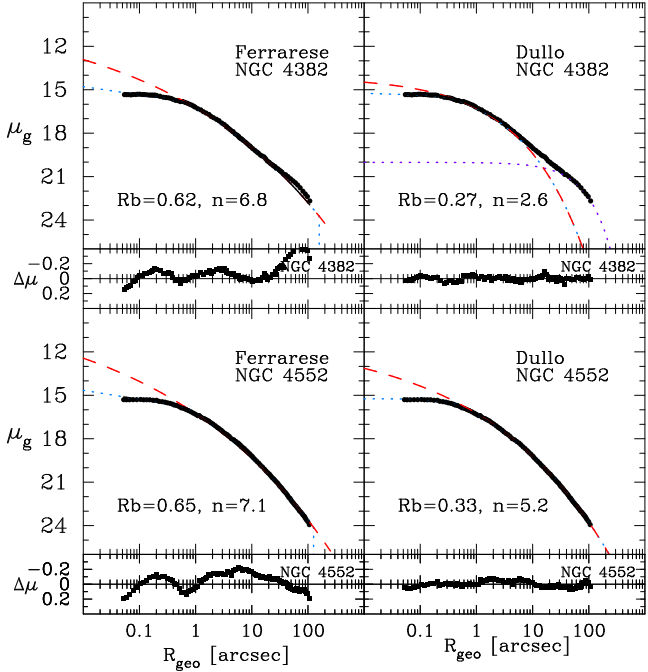}
\caption{
Left panels: Core-S\'ersic models presented in Ferrarese et al. (2006).
Right panels: Our modelling of their published profiles, including an outer
exponential disk (dotted line) for NGC~4382. 
Note: while the innermost data points of the flat cores {\it are} slightly 
affected by the PSF, which we have
not accounted for here and thus things look worse for the Ferrarese et al.\
fit than they are at small radii, this is not the reason for the different 
S\'ersic indices and break radii. 
For the four other core-S\'ersic galaxies that we have in common with
Ferrarese et al. (2006), 
we agree on the location of the break radii (Figure~\ref{Fig4_a}). 
}
\label{Fig4_c} 
\end{figure}

The break radii presented in Trujillo et al.\ (2004) agree with our values for all 3
`core' galaxies that we have in common. 
While 4 of the break radii from the 
6 `core' galaxies in common with Ferrarese et al.\ (2006) agree with
our values, NGC~4382 and NGC~4552 are discrepant (see Figure~\ref{Fig4_a}).  
As noted by Ferrarese et al.\ (2006, their section~5.2), 
the outer disk in the peculiar S0 galaxy NGC~4382 can, if not modelled as a
separate component when the data extends into the domain of the disk, bias (high)
the S\'ersic index that would otherwise be ascribed to the bulge of this galaxy. 
We have therefore modelled their extended light profile for NGC~4382 with a core-S\'ersic
plus outer exponential disk to show this.  Our 
break radius, and S\'ersic index (see Figure~\ref{Fig4_b}), differ from the values presented in Ferrarese et
al.\ (2006) because this galaxy's outer disk did indeed bias their analysis 
(Figure~\ref{Fig4_c}).  
Modelling the light profile from 
Ferrarese et al.\ (2006) for NGC~4552, we recover
their fit when using their published parameters but find that it can be
substantially improved upon with a smaller break radii and a value of 
$\alpha = 2$ (Figures~\ref{Fig17} and \ref{Fig4_c}).  This should be compared
with Kormendy et al.\ (2009, their figure~56) which reports a core 
radius in excess of 1 arcsecond from their visual inspection. Having accounted for NGC 4382 and NGC 4552, Fig.\ 4 reveals an excellent agreement between the break radii obtained by us, Trujillo et al.\ (2004) and Ferrarese et al.\ (2006). 
Marginal discrepancies in the break radii may also arise because of the
different filters used in each study.  While we use data from the F555W
filter, Trujillo et al.\ (2004) primarily used data from a F702W filter, and
we have taken the F475W data, rather than the F850LP data, from Ferrarese et
al.\ (2006) as it most closely matches our F555W data.

When the radial extent of one's data does not adequately probe the curvature
of a spheroid's light profile, one may not recover the correct S\'ersic index. Graham et al.\ (2003) noted that truncating a profile from $\sim$(2-3)$R_{e}$ to $\sim1R_{e}$ changes the fit parameters by up to 5$\%$.
Concerned about this, as our data only extends to 10$\arcsec$, we have additionally compared our S\'ersic indices with
those from Trujillo et al.\ (2004) and Ferrarese et al.\ (2006) which had a larger radial extent. 
First, we note that the S\'ersic index for NGC~4458 from Trujillo et al.\ (2004) was 
biased (high) by the presence of a nuclear disk that was not separately modelled
as we have done here.  They noted that they were not confident in their analysis of
this galaxy and as such we have excluded this one galaxy from Trujillo et 
al.\ (2004) in our Figure~\ref{Fig4_b}. 
With the exception of NGC~4552 and NGC~4382, 
mentioned in the preceding paragraph, the agreement between the S\'ersic 
indices is good to within 50 per cent or better, which
in fair agreement with the 1$\sigma$ uncertainty range of $\pm$36 percent
found by Allen et al.\ (2006, their figure 15).  However, it should also be
noted that due to ellipticity gradients, the major-, minor- and geometric-mean
axis do not have the same S\'ersic index (Ferrari et al.\ 2004). Their values
can disagree by up to a factor of $\sim$2 (e.g.\ Caon et al.\ 1993, their
figure~4).  Consequently, some of the scatter seen in Fig.~\ref{Fig4_b},
which compares our major-axis S\'ersic indices with the geometric-mean axis
values from Ferrarese et al.\ (2006) 
is because of this.

We have also been able to include a comparison of break radii
and S\'ersic indices, in Figs~\ref{Fig4_a} and \ref{Fig4_b}
respectively, for 8 core galaxies that we have in common with Richings
et al.\ (2011).  In general, the agreement is good, although there are
three somewhat discrepant points.
For NGC~5982, we suspect that Richings et al.\ may have missed the
core with their S\'ersic $n=2$ fit
to this large elliptical galaxy with $\sigma = 239$ km s$^{-1}$ and
$M_V = -22$ mag. We are also inclined to
prefer our $n=4$ core-S\'ersic fit to NGC~5813 rather than the $n=9$
core-S\'ersic fit of Richings et al.\ which results in a larger
apparent core.  However we suspect that the dust ring in NGC~3607,
although subtracted, may have still interfered with the optical
light-profile from Lauer et al.\ (2005).  As such, we feel that our
break radius may be over-estimated for this one galaxy, explaining its
outlying nature in the central mass deficit versus black hole mass
diagram (see section~\ref{Sec_Def}).

Recently, Dhar \& Williams (2011) noted that surface brightness
profiles of elliptical galaxies can be modelled well using the
multi-component DW-function (Dhar \& Williams 2010), with smaller rms
residuals than the S\'ersic, Core-S\'ersic and Nuker models. Although
we fit the profiles (from Lauer et al.\ 2005) with limited radial
ranges, for galaxies that we have in common with Dhar \& Williams
(2011) --- NGC 4365, NGC 4382, NGC 4406, NGC 4458, NGC 4472, NGC 4473,
NGC 4478, NGC 4552 and NGC 4649 --- our rms residuals are smaller than
those obtained with the DW-function (Dhar \& Williams 2011, their
Table 1) except for NGC 4458 (Appendix A~Fig.~\ref{Fig17}).

\section{Break radii measurements}\label{Sec_Rb}
Given that the core-S\'ersic model and the Nuker model yield different core sizes, we have included an extended section on the measurement of the break radii via different methods. 

\subsection{The Nuker model}\label{Sec_Rb-Nuk}

As noted previously, the Nuker team (Lauer et al.\ 1995, 2005) 
used a 5-parameter double power-law model for fitting the inner radial
surface brightness profiles of galaxies. 
Dubbed  the ``Nuker law'', it can be written as
\begin{equation}
I(R)=I_{b}2^{(\beta -\gamma )/\alpha }\left(\frac{R}{ R_{b}}\right)^{-\gamma}
\left[1+\left(\frac{R}{R_{b}}\right)^{\alpha}\right]^{(\gamma -\beta)/\alpha },
\end{equation}
where $I_{b}$ is the intensity at the break radius $R_{b}$.
The negative logarithmic slopes for the inner and outer power-law regions are
denoted by $\gamma$ and $\beta$, respectively, while $\alpha$ controls the
sharpness of the transition. $R_{b}$ represents both the radius of maximum
curvature of this {\it model}, and the location where the local gradient of
the model equals $-(\beta + \gamma)/2$. One can readily appreciate how fitting
this model to a larger radial range, and thus to an increasingly steeper
outermost region of what are curved surface brightness profiles, results in larger values of
$\beta$ and thus larger values of $R_b$. When this occurs, the value of $\alpha$ is also reduced, to accommodate an (artificially)
increasingly broad transition.  Having excluded additional nuclear components,
Lauer et al.\ (2005) tabulated their best fitting Nuker model parameters for
the galaxies used in this study, see also Lauer et al.\ (2007a,b), and we refer to those values in some of the following figures.

In their study of the nuclear regions of early type galaxies, Rest et al. (2001) noted that
 for small values of $\alpha$ (i.e.\ broad transitions), 
the Nuker model's parameter
$\gamma$ is rather a representation of the slope of the brightness profile at
radii much smaller than the image resolution limit. They therefore introduced
another parameter, $\gamma^{\prime}$, that was the negative logarithmic slope of the
Nuker model at $R = 0\arcsec.1$, and which was adopted by Lauer et al.\
(2005, 2007b). Graham et al.\ (2003) had however noted that because this
local logarithmic slope $\gamma^{\prime}$ (at $R=0\arcsec.1$) is a distance dependent quantity, galaxies with identical
surface brightness profiles observed at different distances will have
different $\gamma^{\prime}$ values. That is, this quantity is not a physically robust or meaningful quantity to
use. 

\begin{figure}
\includegraphics[angle=270,scale=.650]{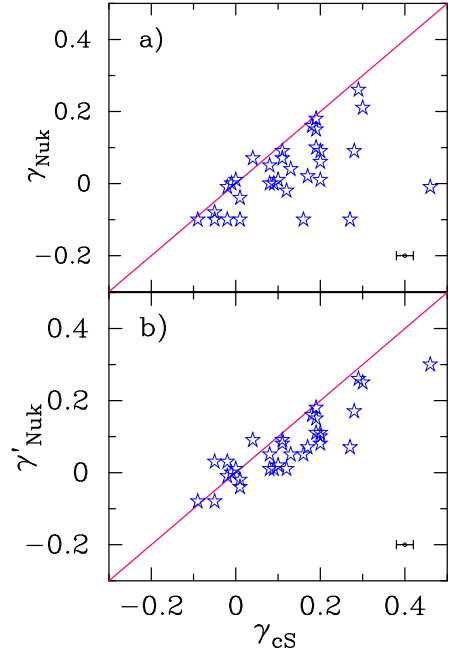}
\caption{
Comparison of the core-S\'ersic model's inner negative logarithmic slopes
($\gamma_{cS}$, Table 2) with a) the Nuker model's inner negative logarithmic
slopes ($\gamma_{Nuk}$, from Lauer et al.\ 2005), and b) the Nuker model's 
local negative logarithmic profile slope at the instrument 
resolution limit ($\gamma^{\prime}_{Nuk}$, from Lauer et al.\ 2005). Representative error bars are shown
at the bottom of each panel.
}
\label{Fig5}
\end{figure}

In Fig.~\ref{Fig5} we show how the $\gamma$ values from the core-S\'ersic
model fit to 32 galaxies (having depleted cores relative to the outer
S\'ersic profile) compare with the $\gamma$ (Pearson's $r=0.53$) and
$\gamma^{\prime}$ (Pearson's $r=0.87$) values obtained from the Nuker model. 
While the latter correlation coefficient is high, it is pointed out that this
does not indicate a slope of unity, which would be required if the two
parameters were equivalent.  Instead, for positive values of $\gamma_{cS}$, 
the plots reveal that the (Nuker model)-derived values of
$\gamma^{\prime}_{\rm Nuk}$ 
and $\gamma_{\rm Nuk}$ are generally smaller than the core-S\'ersic $\gamma$
values. 

\subsection{Model independent approaches}\label{Sec_Rb-ind}

\subsubsection{ $- d^{2}\log~I \, / \, d\log~R^{2}$= \emph{maximum}}\label{Sec_Rb-ind1}

\begin{figure}
\includegraphics[angle=270,scale=0.700]{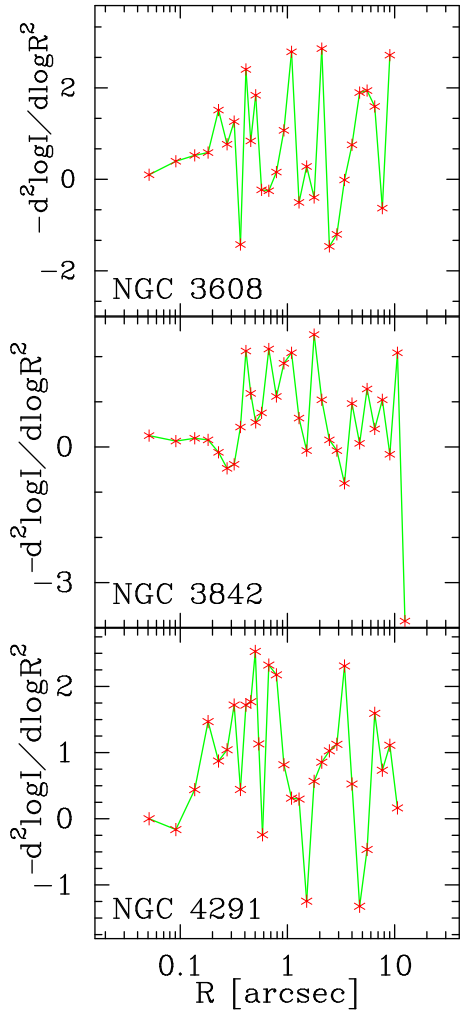}
\caption{Typical examples of the second derivative of the surface brightness
profile (in the\emph{V}-band) plotted against semi-major axis (see
section~\ref{Sec_Rb-ind1}).}
\label{Fig6}
\end{figure}

In an effort to obtain a nonparametric estimate of the break radii, we attempt
to locate the radius corresponding to the maximum of the second derivative of
the observed intensity profile in logarithmic coordinates, 
independent of any model or any smoothing or alteration of the data.
In principle this model-independent radius should mark the break radius. In
practice, however, Fig.~\ref{Fig6} reveals that there exists several comparable maxima
over an extended radial range due to the sensitivity of this approach to the
noise in the profile data.  Therefore, we were unable to use this technique to
acquire accurate break radii.  
We do however show in Fig.~(\ref{Fig7}, Right) that the core-S\'ersic model's
break radius corresponds to the radius where the second logarithmic derivative
of this model has its maximum value. 

In passing we note that Lauer et al.\ (2007a, their figure~17) and Lauer et al.\ (2007b, their
figure~16) had reported a strong consistency between their
Nuker model break radii $R_{b,Nuk}$ and the location of the maximum of the second
derivative of the intensity profile, in logarithmic coordinates, 
acquired from a model-independent approach.  This claim was however surprising because
% as noted before, 
it is well established and understood why $R_{b,Nuk}$ varies considerably as
the fitted radial extent of a galaxy's surface brightness profile is varied
(e.g.\ Graham et al.\ 2003, their figures 2-4).  As such, while the varying $R_{b,Nuk}$ always
corresponds to the radius where the fitted Nuker model has the maximum of its
second derivative, this typically will not correspond to the radius where the
actual data has the maximum in its second derivative.  
Using the same data as Lauer et al.\ (2007a,b), we cannot reproduce their
result.  In Fig.8, we have shown that the second derivative of the observed
galaxy brightness profiles has a large point-to-point variation. It is not
clear how the result of Lauer et al.\ (2007a,b) could be obtained without
smoothing the data.  Furthermore, we continue to reiterate the result of
Graham et al.\ (2003), that the Nuker model fits, especially the break radius,
are dependent on the chosen fitting range. 
In spite of this, that work was the sole basis for their rejection of the
concerns about the Nuker model raised by Graham et al.\ (2003).

\subsubsection{$- d\log~I \, / \, d\log~R =  \gamma^{\prime}=1/2$} \label{Sec_Rb-ind2}
In a continued effort to better measure the sizes of partially depleted galaxy
cores, this section investigates an alternative model-independent
radius that has been used in the literature.  This investigation is important if we are to
accurately quantify the extent of damage caused by coalescing SMBHs at the
centres of galaxies.

Most `core' galaxies have negative, logarithmic, inner profile slopes less
steep than 0.3--0.5 (Glass et al.\ 2011, and references therein), before they
transition to an outer S\'ersic profile with a slope typically steeper than
0.5 (Lauer et al.\ 2005)\footnote{As discussed by Graham \& Guzm\'an (2003),
pure `S\'ersic' galaxies can also have inner profile slopes less steep
than 0.5, and whether or not one measures this (underneath any additional
nuclear components) simply depends on how small one's $R/R_{\rm e}$ spatial
resolution is (see Graham et al.\ 2003, their Figure~6).}.
The negative logarithmic slope beyond the core, in the S\'ersic portion of the
profile, varies as $(b_n/n)(R/R_{\rm e})^{1/n}$.  Graham et al.\ (2003, 
their Figure~6) have revealed that this slope is steeper than 0.5 for S\'ersic
models with $n > 3$--4 once beyond 1\% of the effective radius $R_{e}$.
Given that C\^ot\'e et al.\ (2007) have reported that cores extend to
$\sim$0.02$^{+0.025}_{-0.01}R_{\rm e}$, we can appreciate why the slopes on
the outer side of cores are steeper than 0.5.  As such, the radius where the
negative logarithmic slope of the underlying galaxy surface brightness profile
$\gamma^{\prime}$ equals $ 1/2$ (Carollo et al.\ 1997) can be used to find the 
transition radius between the inner core and the outer S\'ersic profile, and thus
approximate the break radius of the core-S\'ersic model.

\begin{figure}
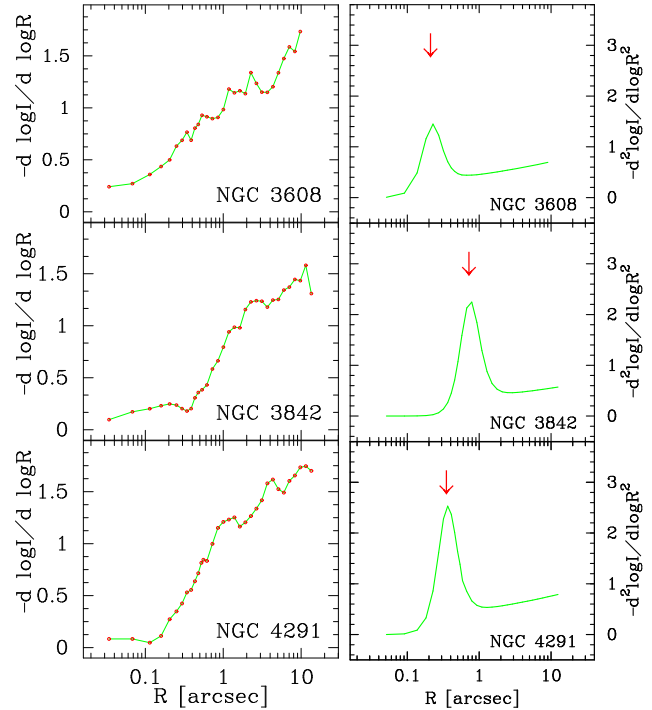

\includegraphics[angle=270,scale=0.490]{nonparslhf.ps} 
\includegraphics[angle=270,scale=.570]{cSmax1.ps}
\caption{Left: Typical examples of the negative logarithmic slopes of the surface
brightness profile (in the\emph{V}-band) plotted against semi-major axis (see
section~\ref{Sec_Rb-ind2}). Right: Typical examples of the second logarithmic derivative of
the fitted core-S\'ersic model (in the 
$V$-band) plotted against semi-major axis. Arrows indicate
the break radii of the galaxies from the fitted core-S\'ersic model
(Table~\ref{Tab2}). }
\label{Fig7}
\end{figure}

\begin{figure}
\includegraphics[angle=270,scale=.60]{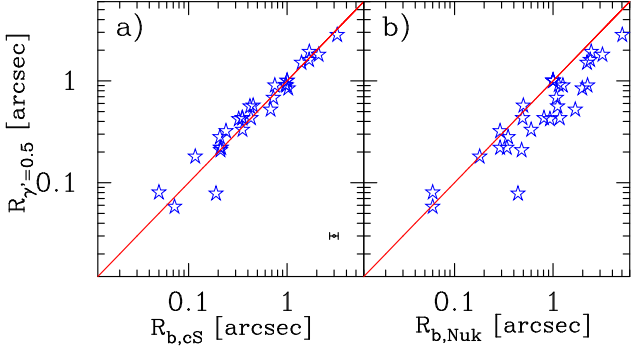}

\caption{
Comparison of $R_{\gamma^{\prime}} = 1/2$ and a) the
break radius from the core-S\'ersic model ($R_{b,cS}$, see Table 2) and b) the
Nuker model ($R_{b,Nuk}$, Lauer et al.\ 2005). 
When present, central light excesses are avoided when determining $R_{\gamma'=1/2}$. NGC 3607, NGC 3640,
NGC 4406, NGC 4589 and NGC 5557 are excluded from this analysis due to
considerable noise in their inner data. A Representative error bar is shown
at the bottom of panel a.} 
\label{Fig8}
\end{figure}

In finding the radius where $\gamma'=1/2$, when present we avoided data points affected by central light excesses before
applying this technique (Fig.~\ref{Fig7}, Left). We visually inspected individual
profiles to double-check these objective break radii estimates and for (only) five galaxies
(NGC 3607, NGC 3640, NGC 4406, NGC 4589 and NGC 5557) the data were too noisy;
as such we excluded them from the comparison of break radii in the following
subsection.  The radii where $\gamma^{\prime} = 1/2$ are plotted in
Fig.~\ref{Fig8} and discussed in the following subsection.

%\begin{center} 
%\end{center}

Before proceeding, we note that Lauer et al.\ (2007a) had also remarked that
Carollo et al.\ (1997) had advocated use of $R_{\gamma^{\prime}}$, with
$\gamma^{\prime} = 1/2$, as a core scale parameter.  They wrote that ``Since
$R_{\gamma^{\prime} = 1/2}$ is generally well interior to $R_{b,Nuk}$, it is
not meant to describe the actual complete extent of the core; it is just a
convenient representative scale.''  To reduce potential misinterpretation of
this comment, 
it is important to note that the Nuker model's $R_{b,Nuk}$ was also never
meant to describe the complete extent of the core. 
While the published Nuker model break radii occur at large radii where the
slope is steeper than $-0.5$ (see Fig.~\ref{Fig8}b), the actual outer edge of
the Nuker model's transition region occurs considerably further out than this, greater than both
 the Nuker model break radii and the outer edge of the
core-S\'ersic model.  Moreover, due to the curved nature of the outer S\'ersic
profile, the outer edge of the Nuker model increases as the fitted radial extent is increased.

\begin{figure}
\includegraphics[angle=270,scale=.56]{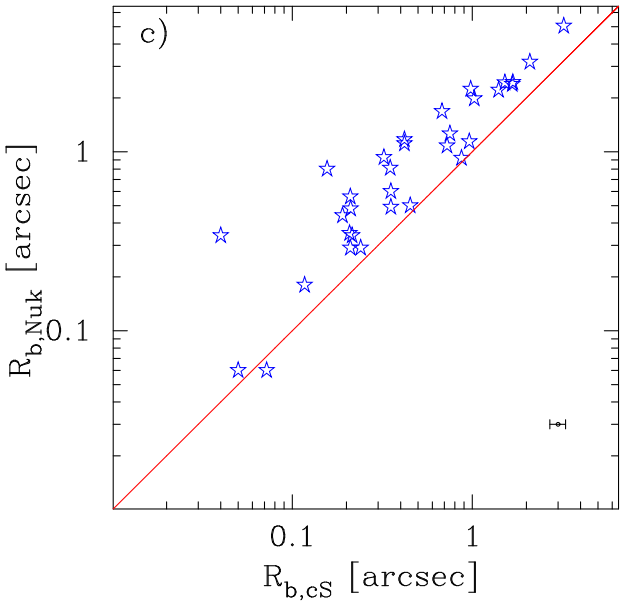}

\caption{Comparison of the published Nuker break radii (Lauer et al.\ 2005) and the core-S\'ersic break radii (see Table 2).  A representative error bar is shown
at the bottom of the panel.}
\label{Fig9}
\end{figure} 

\begin{figure}
\includegraphics[angle=270,scale=.56]{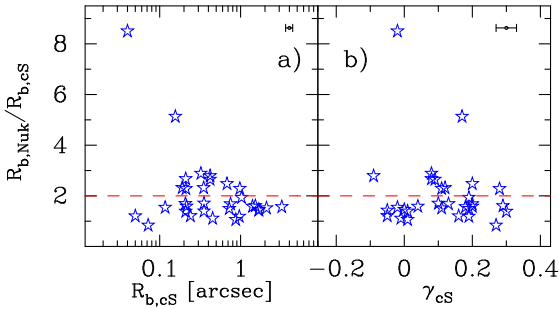}
\caption{The ratio of the Nuker break radius ($R_{b,Nuk}$, Lauer et al.\ 2005) to the core-S\'ersic break radius ($R_{b,cS}$, see Table 2) for core galaxies as a function of a) $R_{b,cS}$ and b) the core-S\'ersic model negative inner logarithmic slope $\gamma_{cS}$.}
\label{FigR12}
\end{figure}

\subsection{The core-S\'ersic model}\label{Sec_Rb-cS}

Fig.~\ref{Fig8}a reveals a strong correlation (Pearson's $r=0.96$) between the break radii
estimated from the model-dependent (core-S\'ersic) and the model-independent
($R _ {\textrm \small{{\gamma^{\prime} = 1/2}}}$) quantitative analysis.  
One can similarly show that the negative logarithmic 
slope of the fitted core-S\'ersic models has a value of 0.5 at a radius very
close to the core-S\'ersic model's break radius. 
This important result has not been noted before.  
Turning things around, this agreement 
endorses the use of $R _ {\textrm
\small{{\gamma^{\prime} = 1/2}}}$ when the data are not too noisy, albeit 
with the caveat that even galaxies without partially depleted cores may still have 
a resolvable radius where $\gamma^{\prime} = 1/2$. 
The core-S\'ersic model is therefore still recommended for identifying and
quantifying cores. 

For comparison, Figs.~\ref{Fig8}b and \ref{Fig9} show that the 
measurement of the Nuker model's break radii presented in Lauer et al. (2007a,b) against the model-independent
break radii $R _ {\textrm \small{{\gamma^{\prime} = 1/2}}}$ ($r=0.91$) and the
core-S\'ersic break radii ($r=0.90$).
Graham et al.\ (2003) explained that increasing the fitted radial range of the
Nuker model, into the curved profile beyond the core, will increase the slope
of the Nuker model's outer power-law $\beta$ and thus result in the Nuker
model's break radius --- corresponding to the location where the model's
slope is the average of the inner and outer power-law slopes $\gamma$ and
$\beta$ --- marching out to larger radii.  For this reason the Nuker model's
break radius can be pulled out beyond the actual transition radius between the core
and the outer S\'ersic profile. On average, the Nuker break radii are $\sim$2 times bigger than the core-S\'ersic break radii (Fig~\ref{FigR12}). While Lauer et al.\ (2007a, their Appendix~C)
refuted this criticism of the Nuker model, they simultaneously reiterated this
very problem and subsequently used $R _ {\textrm \small{{\gamma^{\prime} =
1/2}}}$.  However they did not show how $R_{\gamma'
=1/2}$ compared with the Nuker
model break radii, which is done here for the present galaxy sample 
that have `real' cores (Fig~\ref{Fig8}b).  
The above problem with the Nuker model was, in part, the motivation for the
core-S\'ersic model. 
Fig.~\ref{FigR12} reveals that there is no 
correlation between the ratio of the Nuker break radius ($R_{b,Nuk})$ and the
core-S\'ersic break radius ($R_{b,cS}$) with either a) $R_{b,cS}$ (Pearson coefficient
$r=-0.29$, Fig~\ref{FigR12}a) or b) the core-S\'ersic negative inner
logarithmic slope $\gamma_{cS}$ (Pearson coefficient $r$= -0.12,
Fig.~\ref{FigR12}b).

From Table~\ref{Tab1}, we see that the break radii are smaller than
0.5 kpc.  It would appear that
the 500 pc resolution models by Martizzi et al.\ (2012, their Figure~7)
may have `over-cooked' core-formation in their simulations of galaxies
with cores up to $\sim$8--10 kpc in size.  Similarly, the large
$\sim$3 kpc cores created by Goerdt et al.\ (2010) are not observed in
real galaxies.

\section{Identification of depleted cores, and their slopes}\label{Sec_core}

Initially, Kormendy et al.\ (1994) and Lauer et al.\ (1995) identified cores
if the inner slope of the Nuker model was less than 0.3.  Kormendy (1999)
subsequently relaxed this criteria to read ``galaxies that show a break from
steep outer profiles to shallow inner profiles'', with the ``outer profile''
modelled by the outer power-law of the Nuker model.  However this definition
of a core, and modelling of the stellar distribution, resulted in a
disconnection with the curved outer S\'ersic profile that was known to exist (e.g.\
Caon et al.\ 1993).  Graham et al.\ (2003) therefore advocated that cores be
identified and defined as a central stellar deficit relative to the outer
S\'ersic profile.  Kormendy et al.\ (2009) quoted and partially embraced this
new definition but opted to identify by eye the region to fit the S\'ersic
model and thus the onset of the break radius, rather than using the
core-S\'ersic model in an objective analysis. Their visual core classification agrees with our core identification, however, their approach resulted in
break radii notably larger than those given by the core-S\'ersic model (cf.\
Graham 2004; Ferrarese et al.\ 2006) and thus, given the results in the
previous section, their break radii do not agree with the model-independent
measurement of where the light profile has reached a steep outer profile with
negative logarithmic slope equal to 0.5. This arose in part due to the lack of an infinitely sharp transition region. These larger break radii also result in larger estimates of the central mass deficit. Using a model-independent technique based on average profiles, Hopkins \& Herquist (2010) reported an upper limit to the mass deficit of 2-4 times the central black hole mass, in agreement with the analysis presented in Graham (2004) using the core-S\'ersic model. This is in contrast to the values of 10-20 times the central black hole mass reported by Kormendy \& Bender (2009). 

\begin{figure}
\includegraphics[angle=270,scale=.52]{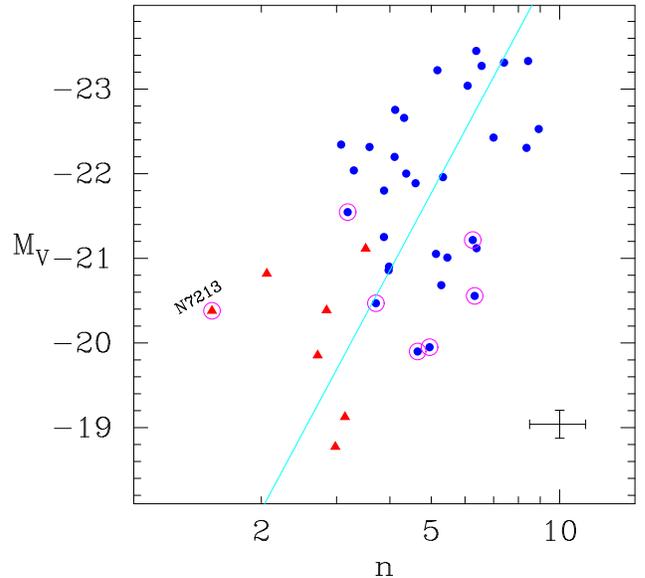}
\caption{
Absolute \emph{V}-band galaxy magnitude from Table 1 (converted to bulge
magnitude for the lenticular and spiral galaxies) plotted as a function of the
S\'ersic index, $n$, that quantifies the shape of the underlying galaxy or
bulge, major-axis light profile. The line M$_{B}=-9.4\log(n)-14.3$ is taken from Graham \&
Guzm\'an (2003, their figure 10) and adjusted here to the $V$-band using
$B-V=0.9$ (Fukugita, Shimasaku \& Ichikawa 1995). Filled circles are the core
galaxies; filled triangles are the S\'ersic galaxies; bulges are circled. The Pearson coefficient between M$_{V}$ and $n$ is $r$= $-0.34$. A representative error bar is shown
at the bottom of the panel, but the 1$\sigma$ uncertainty on the bulge magnitudes for disk galaxies is $\sim$ 0.75 mag.}
\label{Fig10}
\end{figure}

Although all of the 39 galaxies in our sample were tabulated as `core'
galaxies by the published analysis using the Nuker model (Lauer et al.\ 2005,
their Table~4)\footnote{Although tabulated as having a `core' profile in Lauer
et al.\ (2005), those authors are aware that NGC 4486B is not a 
`real' core galaxy (Lauer et al.\ 1996).}, 
in section~\ref{Sec_Fit} we effectively reclassified seven as
S\'ersic galaxies without partially depleted cores. This concern over misidentification in lower luminosity spheroids was first 
highlighted by Graham et al.\ (2003), and the discrepancy is fundamentally due
to the inclusion of S\'erisc galaxies with low S\'ersic index $n$ (and thus a
shallow inner profile slope) that have no depleted core relative to the 
S\'ersic profile which describes the outer galaxy light distribution.  The
typical value of the S\'ersic shape index ($n$) for the seven S\'ersic galaxies is
$\approx 3$ (Fig.~\ref{Fig10}).

\begin{figure}
\includegraphics[angle=270,scale=.90]{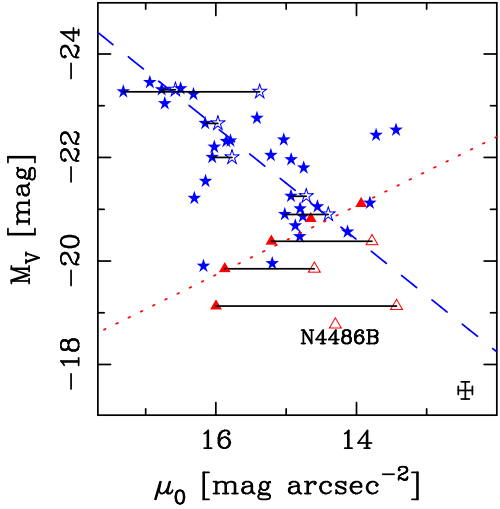}
\caption{Absolute $V$-band magnitude of the galaxy (or the bulge of the 
disk galaxies) plotted against the core-S\'ersic model's inner $V$-band
surface brightness $\mu_{0}$ at the instrument resolution limit 
$R=0\arcsec.0455$ (excluding NGC~1374 and NGC~1399 for which there
were no $F555W$ profiles).  Filled stars and triangles are core-S\'ersic and
S\'ersic galaxies respectively. 
Galaxies with additional nuclear
light have the host galaxy plus central excess flux, at $R=0\arcsec.0455$, 
shown by the open symbols. 
Solid lines connect the central brightness values of the nucleated 
galaxies and their underlying host galaxies. 
The dotted line is taken from Graham \& Guzm\'an (2003, their figure~9) 
and adjusted here to the $V$-band using $B-V = 0.9$. 
The dashed line, $M_{V}=-1.09 (\mu_{0}-16.0)-22.60$, 
shows our least-squares fit to the $M_{V}$ and $\mu_{0}$ relation for galaxies
with `real' depleted cores, and having a Pearson's coefficient $r=-0.44$. A representative error bar is shown
at the bottom of the panel, but the 1$\sigma$ uncertainty on the bulge magnitudes for disk galaxies is $\sim$ 0.75 mag.}
\label{Fig9-10b}
\end{figure}

In Figure~\ref{Fig9-10b} we have plotted the central surface brightness, $\mu_{0, V}$, of the
galaxies or bulges obtained from the fitted models against their $V$-band absolute magnitude $M_{V}$.  As with the $M_V$--$n$
diagram (Fig.~\ref{Fig10}), it is immediately apparent that the seven non-core
galaxies are not some random sample from the 39 galaxies.  They reside in a
region of the $M_V$--$\mu_{0,V}$ diagram known to be occupied 
by galaxies without partially depleted cores.  

Prior to Lauer et al.\ (2005), Trujillo et al.\ (2004) had already revealed
that two of these seven galaxies (NGC~4458 and NGC~4478) had no partially-depleted
core.  Ferrarese et al.\ (2006) additionally identified that NGC~4473 did not
have a depleted core, and 
Kormendy et al.\ (2009) subsequently acknowledged that none of these three
galaxies have depleted cores but instead have an excess of nuclear flux. 
In addition to the double-nucleus galaxy NGC 4486B which is known to have a false core 
(Lauer et al.\ 1996), we find that NGC~1374, NGC~5576 and NGC~7213 also do not have 
cores depleted of stars relative to their host spheroid's (outer) S\'ersic profile.
NGC~7213 is a rather faint spiral galaxy with $M_B \sim -19.5$ mag and is thus
not expected to have a depleted core like luminous, boxy spheroids do.
NGC~1374 is also not a luminous galaxy; it too has $M_B \sim -19.5$ mag, 
and NGC~5576 is only 0.7 mag brighter. 

Besides their distribution in Figs.~\ref{Fig10} and \ref{Fig9-10b}, it is of interest to
examine if there are additional characteristics among these seven galaxies
which have cores according to the Nuker model (Lauer et al.\ 2005) but do not
have partially depleted cores relative to their outer S\'ersic profile.
From the full sample of 39 galaxies, nine have velocity dispersions 
$\sigma \le 183$ km s$^{-1}$ according to HyperLeda, and seven of these nine
galaxies are the above S\'ersic 
galaxies.  Of the remaining two galaxies, NGC~3640 has $\sigma = 182$ km
s$^{-1}$ and a very small, questionable depleted core, and while NGC~4382 has
$\sigma = 179$ km s$^{-1}$ according to HyperLeda's mean value however the most recent
measurement of $\sigma = 205\pm8$ km s$^{-1}$ (Bernardi et al.\ 2002) is 
supportive of a `real' core.

Aside from the above mentioned confirmation by other authors, there are good
reasons to suspect that the Nuker model struggles to identify cores which have
been depleted relative to the inward extrapolation of the outer S\'ersic profile 
in galaxies and bulges fainter than $M_V \approx -21$ mag (see Figure~\ref{Fig10}).
While Lauer et al.\ (2007a) used their Nuker model parameters to determine
where the fitted Nuker model has a slope of $-1/2$ --- which we consider to be a
better measurement of the transition from a shallow core to a steep outer profile ---
one is left with the problem of not knowing which galaxies actually have depleted
cores, that consistent with non-parametric identification methods, relative to the smooth inward extrapolation of the outer S\'ersic
profile.  Figure~\ref{Fig10} reveals that knowledge of this S\'ersic index can
help with this diagnosis, but it obviously requires remodelling the light
profiles.  An additional criteria, which could be better quantified with a larger galaxy
sample, is that alleged cores in galaxies with $\sigma \le 183$ km s$^{-1}$
are likely not to correspond to a real deficit of stars.

\begin{figure}
\includegraphics[angle=270,scale=.50]{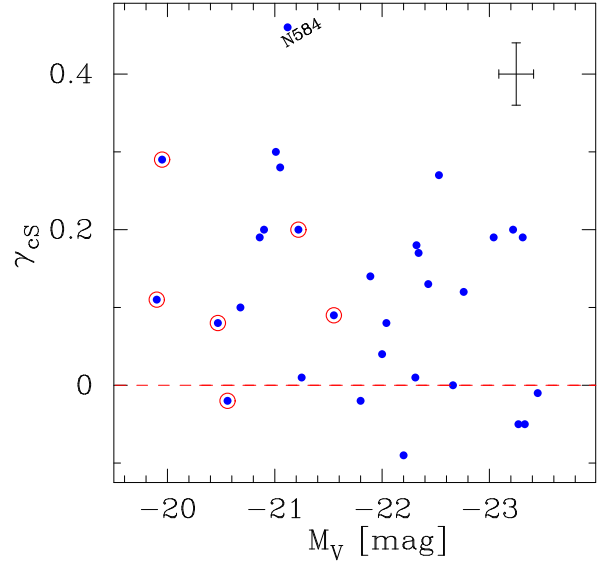}
\caption{Comparison of the core-S\'ersic model's negative inner logarithmic slope
$\gamma$ with the $V$-band absolute galaxy magnitude (bulge magnitude for disc
galaxies).  Bulges in disk galaxies are circled. A representative error bar is shown
at the top of the panel, but the 1$\sigma$ uncertainty on the bulge magnitudes for disk galaxies is $\sim$ 0.75 mag.}
\label{Fig11}
\end{figure}

Although our sample is not suitable for testing the bimodality, or lack
thereof (Glass et al.\ 2011, and references therein), in the distribution of
host galaxy central surface brightness slopes against absolute magnitude
(which, for the reason we have just seen, can not be used as a diagnostic of
core identification), we do find that $\gamma ~ \la 0.3$ for all
(core-S\'ersic)-identified core galaxies, with the possible exception of only NGC 584 (where
$\gamma = 0.46$, see Table 2, and Fig.~\ref{Fig11}).  The low-luminosity
(low-$n$) S\'ersic galaxies without cores can have inner slopes
ranging from 0 to 0.5 and steeper depending on the radius where one samples the underlying host spheroid's S\'ersic profile ( Graham \& Driver 2005, their section 2.4).
Fig.~\ref{Fig11} plots the
relation between $\gamma$ and the $V$-band absolute magnitude $M_{V}$ for the {\it`true'} core galaxies given in Table~2.  
For the first time using the
core-S\'ersic model, we find that a number of bright
(high $n$) core galaxies exhibit values of $\gamma$ less than 0 (see also
Lauer et al.\ 2005, and references therein for galaxies with Nuker model
values of $\gamma < 0$).  Earlier studies with the core-S\'ersic model had
been confined to reporting $\gamma ~ \ga 0$.  Kandrup et
al.\ (2003) have discussed how black hole binaries may couple with the innermost
stars and transport them to a larger radius, resulting in such surface brightness
profiles. 

\section{Structural parameter relations}\label{Sec_rel}

Having selected a set of cores based on the core-S\'ersic model, we intend to
obtain radial profiles over a greater radial extent for those 32 galaxies in a
future paper. Fig.~\ref{Fig10} illustrates the linear correlation between the absolute
galaxy magnitude (bulge magnitude in the case of disk galaxies) and the light
profile shape, $n$. As noted in section 2, the bulge magnitude for the one spiral (Sa) and six 
lenticular (S0) galaxies in our sample are obtained using the representative 
bulge-to-disk flux ratio 
given in Graham \& Worley (2008, their Table 5). 
Although our galaxies are limited in number and range of
absolute magnitude (-18.77 $ > M_{V} > $ -23.45 mag), the overall distribution
in the $L$--$n$ diagram is in
agreement with that from Caon et al.\ (1993), Graham \& Guzm\'an
(2003; their Fig.\ 10) and Ferrarese et al.\ (2006). Given that the S\'ersic index $n$ is derived from $10\arcsec$ profiles, we do
not consider it to be as accurate as possible.
Nonetheless, on average, the galaxy ensemble adhere to the established $L$-$n$
relation, and are in fair agreement with the indices derived from fits to a
larger radial extent (see section 4.1). This gives us some confidence that our S\'ersic
parameters (which are only used once in this paper to derive a `ballpark'
result in section~\ref{Sec_Def}) are not too far off from the correct
values.

\subsection{$R_{b}$-$L$,  $R_{b}$-$\sigma$,  $R_{b}$-$\mu_{b}$, $\mu_{b}$-$L$ and  $\mu_{b}$-$\sigma$ relations}\label{Sec_Rels}

Building on earlier works which explored the connection between early-type
galaxy dynamics and isophotal shape (e.g.\ Davies et al.\ 1983; Nieto \&
Bender 1989; Nieto et al.\ 1991),
% Kormendy \& Bender 1996),  4 references excessive here
Faber et al.\ (1997) highlighted
associations between the core structure and global galaxy properties. Bright
ellipticals with boxy isophotes, slow rotation and pressure supported dynamics
are `core' galaxies, while fainter elliptical galaxies with elliptical or  
`discy' isophotes and often
rotational support are `S\'ersic' (`power-law') galaxies. They went further
to argue that the presence of a `core' is a better predictor of the slow
rotation or boxiness than the galaxy absolute magnitude, although we have just
learned that some of the Nuker-derived cores are not consistent with core-S\'ersic cores or visual classification, i.e. they have been claimed to exist in galaxies with no depleted core.
It would be remiss if we did not 
use our refined core-S\'ersic parameters to derive updated 
scaling relations for galaxies that have central stellar deficits relative to 
their outer S\'ersic profile.  That is, we exclude those galaxies having no depleted
cores.

\begin{figure*}
\includegraphics[angle=270,scale=.80]{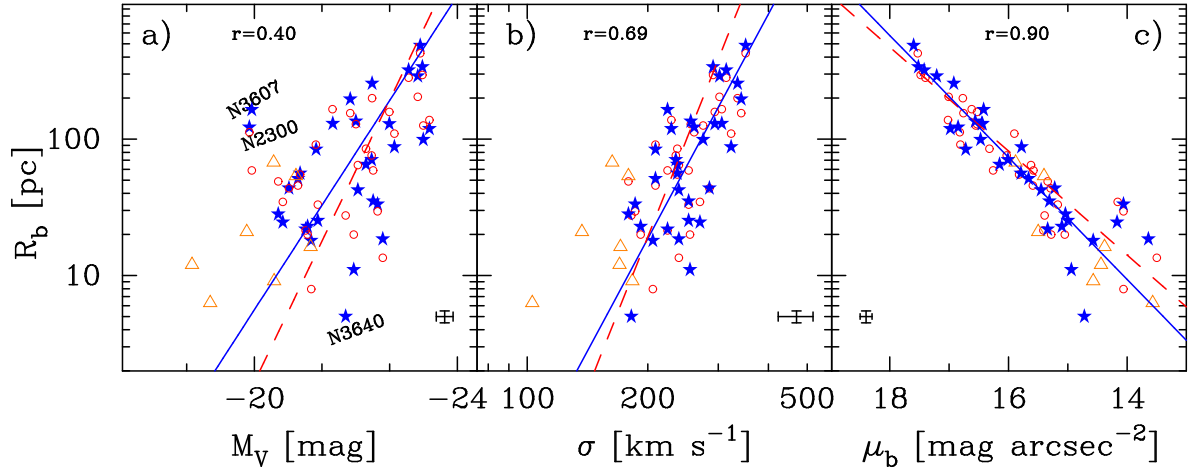}
\caption{
The core-S\'ersic break radius and the published
Nuker `cusp radius' $R_{\gamma'=1/2}$ (Lauer et al.\ 2007a),
collectively $R_{b}$ (in pc), are plotted as a function of a) absolute
$V$-band spheroid magnitude (Table~\ref{Tab1}), b) central velocity dispersion
$\sigma$ (Table~\ref{Tab1}) and c) the associated $V$-band surface
brightness, collectively referred to as $\mu_{b}$ here 
(excluding NGC~1399 for which there was no F555W profile). 
Filled stars are the core galaxies from this study fitted with the
core-S\'ersic model, while open circles are these same galaxies with $R_{b}$
and $\mu_{b}$ ($\gamma^{\prime}=1/2$) 
data from Lauer et al.\ (2007a).  Triangles show the location of
the seven alleged `core' galaxies which are reclassified here as S\'ersic
galaxies and thus have no core-S\'ersic break radii. 
The solid lines are least-squares fits to the core-S\'erisc
data, while the dashed lines are the published Nuker 
relations after they excluded galaxies with $M_{V} > -21$~mag 
(Lauer et al.\ 2007a, their Eqs.\ 13, 14 and 17). Pearson correlation coefficients, $r$, (and representative error bars), for our data, are 
shown at the top (bottom) of each panel. For the disk galaxies, the 1$\sigma$ uncertainty on the bulge magnitudes is $\sim$ 0.75 mag.}
\label{Fig12}
\end{figure*}

Figs.~\ref{Fig12}a and \ref{Fig12}b display the relation between the
core-S\'ersic break radius (Table 2) and the published Nuker `cusp radius' $R_{\gamma'=1/2}$ (Lauer et al.\ 2007a), collectively $R_{b}$, and a) the \emph{V}-band
absolute magnitude $M_{V}$ listed in Table 1, and b) the central velocity
dispersion $\sigma$ (Table 1). Using the ordinary
least-squares (OLS) bisector regression (Feigelson \& Babu 1992), a fit to the core-S\'ersic $R_{b}$ and $\sigma$ yields
\begin{equation}
\mbox{log}\left(\frac{R_{b}}{\mbox{pc}}\right)= (5.47\pm 0.68)~\mbox{log}\left(\frac{\sigma}{200~\mbox{km s} ^{-1}}\right) +~(1.27~ \pm 0.11),~~   
\end{equation}
%200 km s^{-1}
and further application of the bisector regression gives the relation between the core-S\'ersic 
 $R_{b}$ and $M_{V}$ as
\begin{equation}
\mbox{log}\left(\frac{R_{b}}{\mbox{pc}}\right)= (-0.58\pm 0.09) (M_{V}+22) +~(1.90 ~\pm 0.10). ~~~~    
\end{equation}

Similar trends between break radius and galaxy magnitude were also seen in Faber
et al.\ (1997, their figure~4), Ravindranath et al.\ (2001, their
figure~5a,b), Laine et al.\ (2003, their figure~9), Trujillo et al.\ (2004,
their figure~9), de Ruiter et al.\ (2005, their figure~8) and Lauer et al.\
(2007a, their figure~19, bottom panel). The three outliers in our
$R_{b}$-$M_{V}$ relation are: NGC 2300 and NGC 3607--- lenticular (S0)
galaxies with a big core for their bulge brightnesses--- and NGC 3640, a
galaxy known for its morphological peculiarity which probably signals an on
going or a recent merger (Michard \& Prugniel \ 2004).
Shown in Fig.~\ref{Fig12}c is $R_{b}$ (core-S\'ersic break radius from Table 2 plus Nuker `cusp radius' from Lauer et al.\ 2007a) as a function of $\mu
_{b}$ (the \emph {V}-band surface brightness at the core-S\'ersic break radius
 and the Nuker `cusp radius', from Table 2 and Lauer et al.\ 2007a respectively). The bisector fit to the core-S\'ersic $R_{b}$ and $\mu_{b}$ gives
\begin{equation}
\mbox{log}\left(\frac{R_{b}}{\mbox{pc}}\right)= (0.45\pm 0.05)(\mu_{b}-16) +~(1.87~ \pm 0.04),
\end{equation}
or
\begin{equation}
\mu_{b}= (2.24\pm0.28)~\mbox{log}\left(\frac{R_{b}}{100~\mbox{pc}}\right) +~(16.30~ \pm~0.06).
\end{equation}

\begin{figure}
\includegraphics[angle=270,scale=.55]{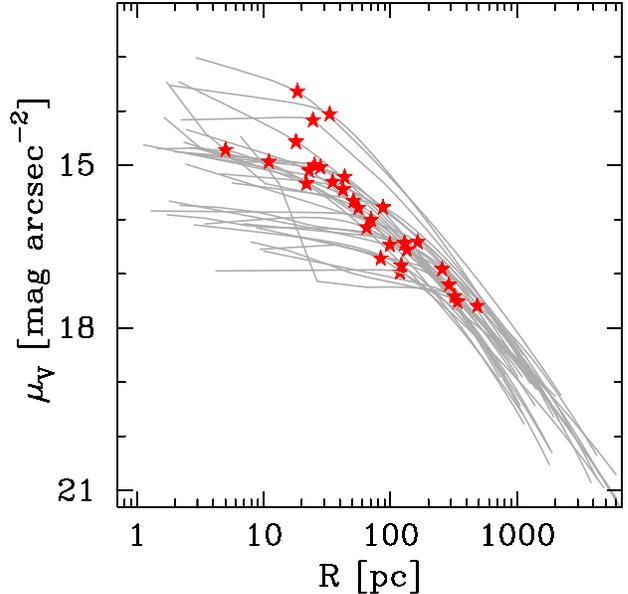}
\caption{A compilation of the core-S\'ersic fits to the major-axis surface
brightness profiles of all core galaxies from Table 2 except for NGC 1399 for
which there was only an 
F606W profile. Stars indicate the break radius of the individual profiles
(cf.\ Fig.~\ref{Fig12}c).}
\label{Fig13}
\end{figure}

\begin{figure*}
\includegraphics[angle=270,scale=.80]{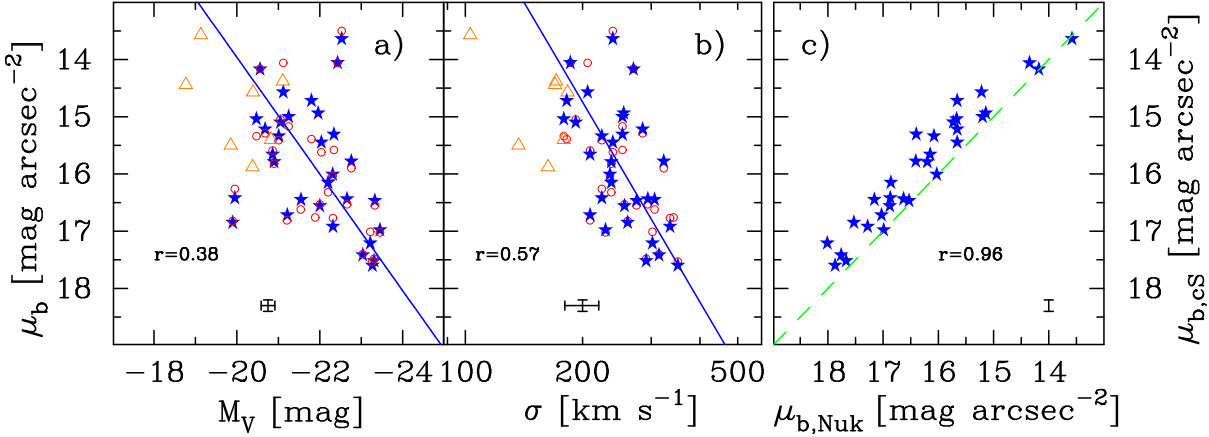}
\caption{
The core-S\'ersic model's break surface brightness
and the published Nuker's `cusp' surface brightness, collectively $\mu_{b}$ 
(\emph{V}-band), are plotted against a) the absolute \emph{V}-band
magnitude of the galaxy (or the bulge for disc galaxies) and b) the
central velocity dispersion $\sigma$ (see Table 1). 
Panel c) plots the
core-S\'ersic $\mu_{b,cS}$ against the published Nuker model's 
estimate of $\mu_{b,Nuk}$ (Lauer et al.\ 2007b).  Filled 
stars are the core galaxies from this study, while open circles are
these same galaxies with $R_{b}$ and $\mu_{b}$ from Lauer et al.\
2007a. Triangles denote the seven Nuker `core' galaxies which are
reclassified here as S\'ersic galaxies. The solid lines (in panels a
and b are Eqs.\ 9 and 10) are the least-squares fits to the
$\mu_{b}$-$L$ and $\mu_{b}$-$\sigma$ core-S\'ersic data. Pearson correlation coefficients, $r$, and representative error bars, for our data, are 
shown at the bottom each panel. For the disk galaxies, the 1$\sigma$ uncertainty on the bulge magnitudes is $\sim$ 0.75 mag.}
\label{Fig14}
\end{figure*} 

Equations 5, 6 and 8 update the $R_{b}$-$\sigma$, $R_{b}$-$L$ and
$\mu_{b}$-$R_{b}$ relations presented in Lauer et al.\ (2007a, their equations
13, 14 and 17 respectively).
Fig.~\ref{Fig13} and Fig.~\ref{Fig12}c confirm, but re-define the tight
correlation between the core-S\'ersic core brightness $\mu_{b}$ and the core-S\'ersic core radius $R_{b}$
(Faber et al.\footnote{Faber et al.\ (1997) ascribed the tight correlation
among the central properties of early-type galaxies to the presence of a ``core
fundamental plane" ($\log~r_{b}$, $\mu_{b}$, $\log~\sigma$), which is analogous
to the global Fundamental Plane --- $\log r_{e}, \mu_{e}, \log~\sigma$ (e.g.\
Djorgovski \& Davis 1987).} 1997; de Ruiter et al.\ 2005, their figure~7;
Lauer et al.\ 2007a, their figure~6). Apparent in Fig.~\ref{Fig13} is a
roughly universal profile beyond the core of `core' galaxies, out to $\sim$1
kpc, which explains the tight relation seen in Fig.~\ref{Fig12}c.

Fig.~\ref{Fig14}c plots the core-S\'ersic $\mu_{b} $ versus the Nuker
$\mu_{b}$. In over-estimating the core radii, where $\gamma^{\prime} \approx 1/2$,
the Nuker model underestimates the associated surface brightness, by typically
1 mag arcsec$^{-2}$ and up to 2 mag arcsec$^{-2}$ with respect to the core-S\'ersic model. In Figs.~\ref{Fig14}a and
\ref{Fig14}b we show relations involving $\mu_{b} $ (the \emph {V}-band surface brightness at the core-S\'ersic break radius
 and the Nuker `cusp radius', from Table 2 and Lauer et al.\ 2007a respectively) with $M_{V}$ and
$\sigma$. Akin to Fig.~\ref{Fig12}a, the (core-S\'ersic)-identified core galaxies display a
correlation between the core-S\'ersic $\mu_{b} $ and $M_{V}$, such that the OLS bisector
regression analysis gives
\begin{equation}
\mu_{b}= (-1.02\pm 0.10) (M_{V}+22) +~(16.00~ \pm~0.20).
\end{equation}
The OLS bisector fit to the core-S\'ersic $\mu_{b}$ and $\sigma$ yields
\begin{equation}
\mu_{b}= (11.61\pm 1.60)~\mbox{log}\left(\frac{\sigma}{200~\mbox{km s} ^{-1}}\right) +~(14.75~ \pm 0.24).
\end{equation}

\subsection{Core size versus black hole mass}\label{Sec_Rb-BH}

\begin{figure}
\includegraphics[angle=270,scale=.90]{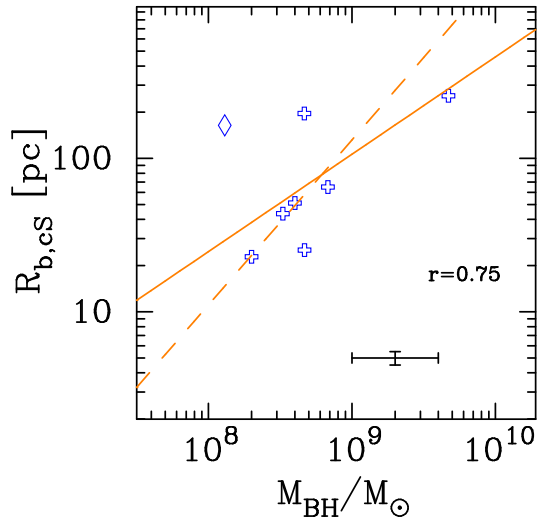}
\caption{
The core-S\'ersic break radius $R_{b,cS}$ (Table 2) plotted as a function of
the black hole mass $M_{BH}$ for 8 core galaxies with direct black hole mass
measurements, obtained from Graham 2008b and Graham et al.\ 2011 and adjusted
to distances given in Table 1. The solid line (Eq.\ 11) is a least-squares fit to all 8
core galaxies, while the dashed line (Eq.\ 12) is a least-squares fit to the 7
core elliptical galaxies (open crosses) after excluding the only
lenticular galaxy NGC 3607 (open diamond) having a Pearson correlation coefficient r=0.75 as shown in the plot. A representative error bar is shown
at the bottom of the panel (\emph{see text for details}).}
\label{Fig15}
\end{figure}

Although the relation between the break radius and the supermassive black hole
(SMBH) mass is less fundamental than the relation between the central mass
deficit and the SMBH mass (e.g.\ Graham 2004; Ferrarese et al.\ 2006, Merritt
2006), de Ruiter et al.\ (2005) and Lauer et al.\ (2007a) nonetheless argue
for the existence of a good correlation between the former. Shown in Fig.~\ref{Fig15}
is the core-S\'ersic break radius plotted against the black hole mass $M_{BH}$
for 8 core galaxies with direct black hole mass measurements (see Graham
2008b and Graham et al.\ 2011).  The bisector fit to $R_{b}$ and $M_{BH}$  for 
these 8 galaxies gives
\begin{equation}
\mbox{log}\left(\frac{R_{b}}{\mbox{pc}}\right)= (0.63 \pm 1.73)~\mbox{log}\left(\frac{M_{BH}}{10^{9}~M_{\sun}}\right) +~( 2.03~ \pm 0.78).~~~~~ 
\end{equation}
However, while considering only 7 of the 8 core galaxies, after excluding the
only disk galaxy (NGC 3607), the regression analysis of $R_{b}$ and $M_{BH}$
yields the more certain relation 
\begin{equation}
\mbox{log}\left(\frac{R_{b}}{\mbox{pc}}\right)= (1.01 \pm 0.69)~\mbox{log}\left(\frac{M_{BH}}{10^{9}~M_{\sun}}\right) +~( 2.07~ \pm 0.33).~~~~~~ 
\end{equation}
Given the two widely used black hole mass estimators, the $M_{BH}$-$\sigma$
relation (Ferrarese \& Merritt 2000; Gebhardt et al.\ 2000) and the
$M_{BH}$-$L$ relation (Marconi \& Hunt 2003), we can construct the
$R_{b}$-$M_{BH}$ relation from the $R_{b}$-$\sigma$ and $R_{b}$-$L$ relations
in section~\ref{Sec_Rels} to further investigate the core size and black hole
connection. Combining the $M_{BH}$-$\sigma$ relation from Graham et al.\
(2011, their final entry in Table 2, acquired using only elliptical galaxies)
with the $R_{b}$-$\sigma$ relation (Eq.\ 5) we can derive the new 
$R_{b}$-$M_{BH}$ relation 
\begin{equation}
\mbox{log}\left(\frac{R_{b}}{\mbox{pc}}\right)= (1.03\pm~0.20)~\mbox{log}\left(\frac{M_{BH}}{10^{9}~M_{\sun}}\right) + (2.08~ \pm 0.22),~~~~  
\end{equation}
 which is in remarkable agreement with Eq.\ 12.
Combining the $R_{b}$-$L$ relation (Eq.\ 6) with the $M_{BH}$-$L$ relation for predominantly massive spheroids 
from Graham (2007, his equation 19) converted to the $V$-band using
$B-V = 0.9$ (Fukugita, Shimasaku \& Ichikawa 1995), yields
 \begin{equation}
\mbox{log}\left(\frac{R_{b}}{\mbox{pc}}\right)= (1.45\pm~0.29)~\mbox{log}\left(\frac{M_{BH}}{10^{9}~M_{\sun}}\right) + (2.03~ \pm 0.16).
\end{equation}

While the $M_{bh}$-$L$ relation is likely to be non-(log-linear), Graham 2012b, the log-linear relation presented by Graham (2007) is dominated by 
massive spheroids and thus a good representation of the $M_{BH}$-$L$ relation for the `core' galaxies. 

It is worth noting that the scatter in the direct $R_{b}$-$M_{BH}$ relation
established using the 7 elliptical core galaxies with direct black hole mass measurements
(Eq.\ 12) is large.  Although we only have a limited
number of galaxies with a measured black hole mass and Eq.\ 12 is somewhat
driven by the highest mass black hole\footnote{Removing the highest mass black
hole and excluding the disk galaxy gives a slope of 1.63$\pm$0.75.}, 
Eq.\ 12 is consistent (over-lapping error bars) with the two inferred
relations (Eqs~13 and 14).  Moreover, in contrast to the discussion in Lauer
et al.\ (2007a, their Eqs~20 and 21), there is a good consistency among the
deduced relations (Eqs.\ 13 and 14). The updated relations are not (i)
contaminated by the inclusion of galaxies without cores nor (ii) based on
galaxy rather than bulge luminosity for the disk galaxies (an issue discussed
by Graham 2008a, his section~6), and (iii) use core-S\'ersic break radii.

\subsection{Central mass deficit}\label{Sec_Def}

Galaxy merging is believed to be a common occurrence, responsible for the
morphologies of elliptical galaxies. Growing evidence, based on extensive
numerical experiment has indicated that the merger remnant of collisions
between nearly equal mass spiral galaxies resemble early-type galaxies (e.g.\
Toomre \& Toomre 1972; Hernquist 1993; Somerville \& Primack 1999; Cox et al.\
2006; Naab et al.\ 2006; Naab \& Ostriker 2009). `Dry' mergers of elliptical
galaxies may subsequently build `core' galaxies (e.g.\ Faber et al.\ 1997;
Khochfer \& Burkert 2003).

In `core' galaxies, the inner most stars are thought to have been ejected by
inwardly spiraling binary SMBHs in the course of such dry, i.e.  dissipationless, galaxy
mergers, producing the observed central luminosity deficit, $L_{def}$, relative to the inward
extrapolation of the outer S\'ersic profile. Multiplying this deficit by the appropriate stellar mass-to-light ratio gives the central mass deficit, $M_{def}$. More specifically, scattering of
stars from the galactic nuclei in three-body interactions (since there is no
significant amount of gas which may render dynamical friction) is thought to
be the avenue through which the coalescing binary black holes will `harden'
and be delivered to the galaxy center (e.g.\ Begelman, Blandford \& Rees 1980;
Nakano \& Makino 1999; Milosavljevi\'c \& Merritt 2001).  Using N-body
simulations, Merritt (2006) showed that core formation is a cumulative
process, where the total central mass deficit after $N$ dry mergers is such
that $M_{def}\approx 0.5 N M_{BH}$. Gualandris \& Merritt (2011) further
discuss the long-term evolution of black hole binaries affecting the central
stellar distribution, and scouring out cores having radii bigger than the
influence of the binaries.

Here we have a) identified the galaxies with scoured out `cores'
(Table~\ref{Tab2}) and b) quantified how their sizes scale with the final
black hole mass $M_{BH}$ (section 7.2). Graham's (2004) estimation of the
central stellar mass deficit ($M_{def}$) through the employment of the
core-S\'ersic model yielded $M_{def} \approx M_{BH}$, a result later confirmed
by Ferrarese et al.\ (2006). This mass deficit is in accord with hierarchical
galaxy formation models, where luminous galaxies are a by-product of about 1
to 2 major dissipationless merger. For reference, based on observations of
galaxy pairs, Xu et al.\ (2011) found that since $z=1$, massive galaxies have
experienced 0.4 to 1.2 major-mergers (see also Bell et al.\ 2004).

\begin{figure}
\includegraphics[angle=270,scale=.550]{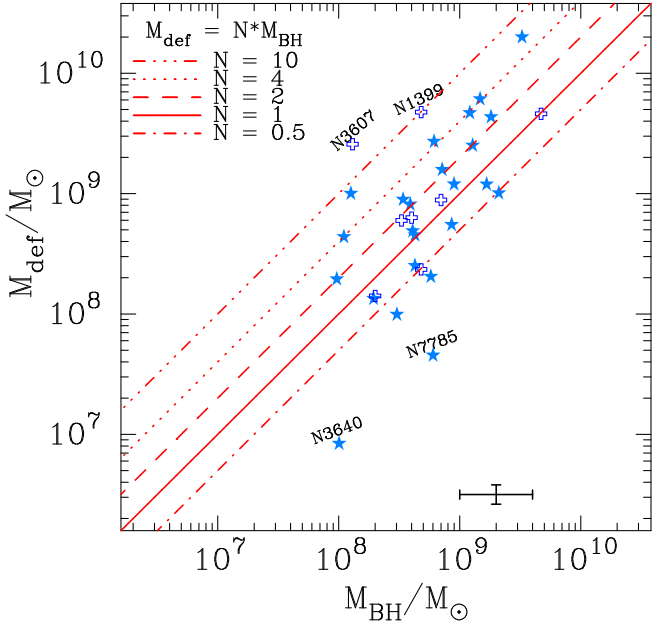}
\caption{Tentative central mass deficit $(M_{def})$ versus black hole mass
($M_{BH}$) for 32 core galaxies. We used Graham (2008b) and Graham et al.\
(2011) for direct supermassive black hole (SMBH) mass measurements of 8 core
galaxies (open crosses), while the $M$-$\sigma$ relation presented in Graham et al.\ (2011)
was used for estimating the SMBH masses of the remaining 24 core galaxies
(filled stars). A Representative error bar is shown
at the bottom of the panel.}
\label{Fig16}
\end{figure}

In Fig.~\ref{Fig16} we present tentative central stellar mass deficits
obtained using our core-S\'ersic model parameters  and Eq.\ A19 from Trujillo et al.\ (2004) plotted against the
observed or predicted (Graham et al.\ 2011) 
supermassive black hole mass for each core galaxy. While
$R_{b}$, $\mu_{b}$ and $\gamma$ are well constrained from our core-S\'ersic
fits, $R_{e}$ and $n$ may be less well constrained (section~\ref{Sec_Lit}) and
as such we caution about over-interpreting the results in Fig.~\ref{Fig16}.
In passing we again note that the break radius and mass deficit for
NGC~3607 may be too large, based on {\it HST/NIC2} data analysed by
Richings et al.\ (2011). Here we simply remark that the mass deficits scatter around 0.5 to 4 times the
central supermassive black hole mass. 
$M_{def}/M_{BH}$ ratios less than 0.5 imply that a minor merger event may have
taken place (in the absence of loss cone refilling). Note that we follow Graham (2004) and assume a $V$-band stellar mass-to-light ratio of $\sim$ 3.5 (Worthey 1994) to compute the mass deficits.
For further reference, using N-body simulations, Kulkarni \& Loeb
(2011), measure mass deficits which are up to 5 times the mass of the
central black hole.

Lauer et al.\ (2007a) proposed an alternate method to quantify the central
stellar mass deficit in terms of a core with zero transition region
breaking to an outer power-law profile with negative, logarithmic
slope $\beta$. However, as revealed by Graham et al.\ (2003), the
Nuker model $\beta$ is not a robust quantity but varies with the
radial range of the light profile that one tries to model.  
The range of (mass deficit)-to-(black hole mass) 
values in Fig.~\ref{Fig16} is notably less than the values in Lauer et
al.\ (2007a) which were as high as $\sim 19$ at $M_{BH} = 10^9 M_{\odot}$ (for
black hole masses predicted using their $M_{BH}$-$\sigma$
relation). Kormendy et al.\ (2009, their figure~42) also reported central mass deficits
that were some $\sim$5 to 20 times greater than the central black hole mass, 
raising further doubts over their method of analyzing the galaxy light profiles (see the model-independent analysis by Hopkins \& Hernquist 2010).

Finally, using Nuker model parameters, G\"ultekin et al.\ (2011) report a
break radius of $0\arcsec.93$ for NGC~4382 and a central mass deficit of
$5.9\times 10^8 M_{\odot}$.
Our analysis yields a break radius
roughly three times smaller ($0\arcsec.32$) and a central mass deficit which
is also three times smaller ($1.8\times 10^8 M_{\odot}$). 

Having refined the core galaxy sample in this study, in a follow-up paper we
intend to acquire a greater radial range of the light-profiles for these
galaxies, enabling a better estimate of the outer S\'ersic parameters and thus
the central mass deficits. This will allow us to check for a positive $M_{def}/M_{BH}$ correlation with host spheroid mass and $M_{BH}$, tentatively seen in Fig.~\ref{Fig16}.

\begin{center}
\begin{table*}
\begin {minipage}{160mm}
~~~~~~~~~~~\caption{Comparison of the detection of nuclear components from different studies}
\label{Tab3}
~~~~~~~~~~~~~~~\begin{tabular}{@{}llccccccccccc@{}}
\hline
\hline
&& Rest et al.\ &Ravindranath & Lauer et al.\ &C\^ot\'e et al.\ &Our result \\
Galaxy &Profile Type &(2001)&et al.\ (2001)&(2005)&(2006)&\\

\multicolumn{6}{c}{} \\ 
\hline                             
NGC 0741&\emph{c-S}& ---&---&Yes&---&Yes\\
NGC 1374&\emph{S}&---&---&No&---&Yes\\
NGC 1399&\emph{c-S}&---&---&Yes&---&No\\
NGC 4278&\emph{c-S}&---&Yes&Yes&---&Yes\\
NGC 4365&\emph{c-S}&No&---&Yes&Yes&Yes\\
NGC 4406&\emph{c-S}&---&No&Yes& No & No\\
NGC 4458&\emph{S}&---&---&No&Yes&Yes\\
NGC 4472&\emph{c-S}&---&No&Yes&No&Yes\\
NGC 4478&\emph{S}&No&&Yes&Possibly&Yes\\
NGC 4486B&\emph{S}&---&---&No&possibly&Possibly\\
NGC 4552&\emph{c-S}&---&---&Yes&No&Yes\\
NGC 5419&\emph{c-S}&---&---&Yes&---&Yes\\
NGC 6876&\emph{c-S}&---&---&No&---&Possibly\\
NGC 7213&\emph{S}&---&---&Yes&---&Yes\\

\hline
\end{tabular} 

\end {minipage}
\end{table*}
\end{center}

\section{Additional nuclear components}\label{Sec_Add}
Additional nuclear light
is detected in 12 ($\sim $31\%) of the 39 galaxies. Table 3 presents a comparison
of detections of additional nuclear components from different studies; as done
by C\^ot\'e et al.\ (2006, their Table 3). Although the majority of our sample
are core galaxies ($\sim$82\%), 5 of the 12 nucleated galaxies are S\'ersic
galaxies. We find good agreement with the work of C\^ot\'e et al.\ (2006) in
assigning the additional nuclear components with only two exceptions (NGC 4472
and NGC 4552). 
NGC~4552 has a point-source AGN (Renzini et al.\ 1995; Carollo et al.\ 1997;
Cappellari et al.\ 1999) with a radio flux $\sim$103 mJy at 1.4 GHz (Condon et
al.\ 1998).  As for
the Sy2 galaxy NGC 4472, at odds with Lauer et al.\ (2005) but in agreement
with C\^ot\'e et al.\ (2006), Ravindranath et al. (2001) also did not detect
any additional nuclear component (Soldatenkov, Vikhlinin \& Pavlinsky 2003).
We do however note that this galaxy's apparent point-source (which has a radio
flux $\sim$752 mJy at 1.4 GHz: White \& Becker 1992) is very faint and the
presence of dust can lead to such small irregularities in profiles acquired
after image deconvolution.

Lauer et al.\ (2005) identified the nuclear light excess as a central upward
deviation from the best Nuker model fit to the host galaxy surface brightness
profile. Their identification is not always consistent with our analysis.  The
relative faintness and extended nature of the additional nuclear components in
NGC 1374 and NGC 4458 are viable explanations as to why
our detections are at odds with Lauer et al.\ (2005). We also do not detect
additional nuclear light in our data for NGC 1399 nor NGC 4406. Ravindranath
et al.\ (2001) and C\^ot\'e et al.\ (2006) also noted the absence of central
light excess in NGC 4406, while Gebhardt et al.\ (2007) have also reported
that NGC 1399 lacks any second component. Lyubenova et al.\ (2008) do however
provide tentative evidence for the existence of a nuclear star cluster (or
swallowed globular cluster) in NGC~1399. Lastly, Rest et al.\ (2001), after
adopting a conservative central nuclei assigning approach reported the absence
of additional nuclear components in NGC 4365 and NGC 4478, while all
successive studies, including ours, identified central light excess in these
two galaxies.

Of the 5 S\'ersic galaxies with additional nuclear component, only one, NGC
7213 (V\'eron-Cetty \& V\'eron 1988), has a central light excess associated
with nonthermal emission from an AGN. We further note that the remaining 4
nucleated S\'ersic galaxies have a nuclear disk, non-AGN central nuclei or
both. In contrast, we note the presence of AGN in (at least) 5 of the 7 nucleated core
galaxies: NGC 741 (Condon et al.\ 2002); NGC 4278 (Younes et al.\ 2010); NGC
4472 (Diehl \& Statler et al.\ 2008); NGC 4552 (Carollo et al.\ 1997;
Cappellari et al.\ 1999) and NGC 5419 (Capetti \& Balmaverde 2005). 
Such prevalence is in accord with the studies by Balmaverde \& Capetti (2006)
and Richings et al.\ (2011), see also Pellegrini (2010). 
The central light excess in the remaining 2 core galaxies is not AGN related: NGC
4365 has a stellar cluster (Carollo et al.\ 1997) and NGC 6876 has a 
double optical nucleus possibly from an inclined disk (Lauer et al.\ 2002). 

\section{Conclusions}\label{Sec_Con}

We have re-modelled the major-axis, surface brightness profiles of 39 alleged
`core' galaxies from Lauer et al.\ (2005), using S\'ersic and core-S\'ersic
models. We have additionally and simultaneously accounted for the point
sources and additional nuclear components that were excluded by the Nuker analysis. Consistent
with earlier published works, we found that the S\'ersic and core-S\'ersic
models yield a robust representation of the underlying light distributions
of S\'ersic and core-S\'ersic galaxies, respectively, all the way to the
resolution limit. The typical rms residual scatter is 0.02 mag arcsec$^{-2}$.

The main results of this work are:\\

1. We have identified 7 of the 39 `core' galaxies from Lauer et al.\ (2005)
   to be S\'ersic galaxies which do not have partially depleted cores relative
   to the inward extrapolation of their outer S\'ersic light profile. 
   This situation tends to arise in galaxies and bulges 
   fainter than $M_V \approx -21$ mag.  Such galaxies with spheroid S\'ersic
   index $n ~ \la 3$ or velocity dispersion $\sigma ~ \la 183$ km s$^{-1}$ are
   not likely to have partially depleted stellar cores. 

2. We provide physical parameters ($R_b, \mu_b, \gamma$) for the cores of 32
   `core' galaxies , derived using the core-S\'ersic model.  

3. Due to noise or real small scale structure, non-parametric core size
   estimations obtained by locating the maximum of the second logarithmic derivative of
   the (non-smoothed) light profile, i.e.\ the point of greatest 
   curvature, appear to be unreliable (Fig.~\ref{Fig6}). 
   
 4. As with the Nuker model, the break radius of the core-S\'ersic model is shown to coincide with the radius where
  it has a maximum in the second logarithmic derivative (Fig.\ 9, Right). 
 
 5. For the first time, the radius where the negative logarithmic slope of the light profile
   $\gamma'=1/2$, considered to be a suitable estimator for the size of the
   core, is shown to be consistent with the core-S\'ersic model break radius
   (Fig.~\ref{Fig8}). 
   It should, however, be noted that even galaxies without depleted cores
   will have a radius where $\gamma'$ equals 1/2. Therefore, this measurement cannot be used to 
   identify `true' depleted-core radii. 

6. We have compared the core-S\'ersic break radii with the Nuker break
   radii. In line with previous works, we found that the Nuker break radii are larger than the core-S\'ersic break radii and also $R_{\gamma'=1/2}$: on average, the Nuker break
   radii are $\sim$2 times bigger than the core-S\'ersic break
   radii. Furthermore, the surface brightnesses ($\mu_{b}$) at the Nuker
   model's break radii are up to 2 mag arcsec$^{-2}$ fainter than the surface
   brightness at the core-S\'ersic model break radii.

7. We have updated various structural parameter relations after excluding
   galaxies which do not have `real' cores, and using core-S\'ersic
   parameters. We have also used the bulge magnitude instead of the galaxy
   magnitude for  the disk galaxies. We provide updated $R_{b}$-$L$, $R_{b}$-$\sigma$,  $R_{b}$-$\mu_{b}$,
    $\mu_{b}$-$L$ and $\mu_{b}$-$\sigma$
   relations in section 7.1. 

8. In contrast to Lauer et al.\ (2007a), we found consistency among three
    linear $R_{b}$-$M_{BH}$ relationships (section~\ref{Sec_Rb-BH}). While one
    of these is obtained directly from $R_{b}$ and $M_{BH}$ data (Eq.\ 12), the
    other two are constructed by combining the $R_{b}$-$\sigma$ and
    $M_{BH}$-$\sigma$ relations and the $R_{b}$-$L$ and $M_{BH}$-$L$
    relations.

9. We detected additional nuclear light in 12 of the 39 sample galaxies. While
    our sample is rich in `core' galaxies (32/39), 5 of the 12 
    nucleated galaxies are S\'ersic galaxies: 1 with nonthermal emission from
    an AGN and 4 with excess stellar light. 
    Five of the 7 nucleated `core' galaxies have AGN emission.
    These results are in good agreement with previous
    estimates (e.g.\ Rest et al.\ 2001; C\^ot\'e et al.\ 2006).

10. Following Graham (2004), we derived a tentative central mass deficit for
   our `core' galaxies using Eq.\ A19 from Trujillo et al.\ (2004). These
   deficits are about 0.5 to 4 times the expected central supermassive black
   hole mass.

\section{Acknowledgments}

This research was supported under the Australian Research Council's
funding scheme (DP110103509 and FT110100263). This research has made use of the
NASA/IPAC Extragalactic Database (NED) which is operated by the Jet Propulsion
Laboratory, California Institute of Technology, under contract with the
National Aeronautics and Space Administration.

\section{Appendix A}
\begin{center}
\begin{figure*}
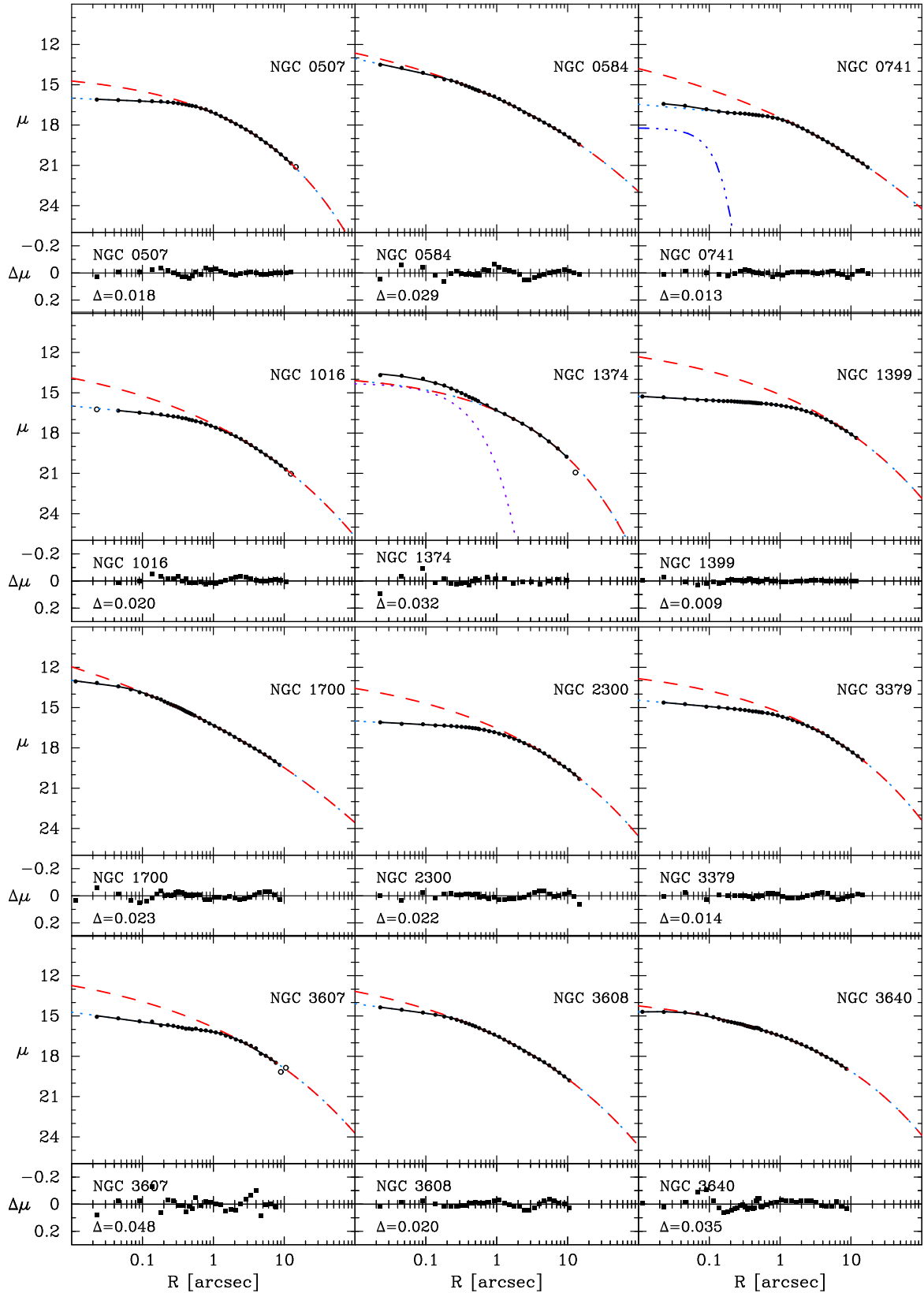

\includegraphics[angle=270, scale=.64]{grp1.ps}
\includegraphics[angle=270, scale=.64]{grp2.ps}
\caption{
Major-axis surface brightness profiles for the galaxies listed in Table 1.
All profiles were obtained with the F555W ($\sim V$-band) filter, except for
NFC~1374 and NGC~1399 which were obtained with the F606W ($\sim R$-band)
filter.  The dashed curves show the S\'ersic component of the core-S\'ersic
fits to the data, while additional nuclear sources were fit with either a
Gaussian (dash-dot-dot-dot curve) or an exponential function (dotted curve).
The solid curves show the complete fit to the profiles, with the rms
residuals, $\Delta$, about each fit given in the lower panels. Data points
excluded from the fits are shown by the open circles.
Brief detail on NGC 1374; NGC 3607; NGC 3640 is provided in
Appendix B, see also the text.}
\label{Fig17}
\end{figure*}
\end{center}

\setcounter{figure}{21}
\begin{center}
\begin{figure*}
\includegraphics[angle=270, scale=.68]{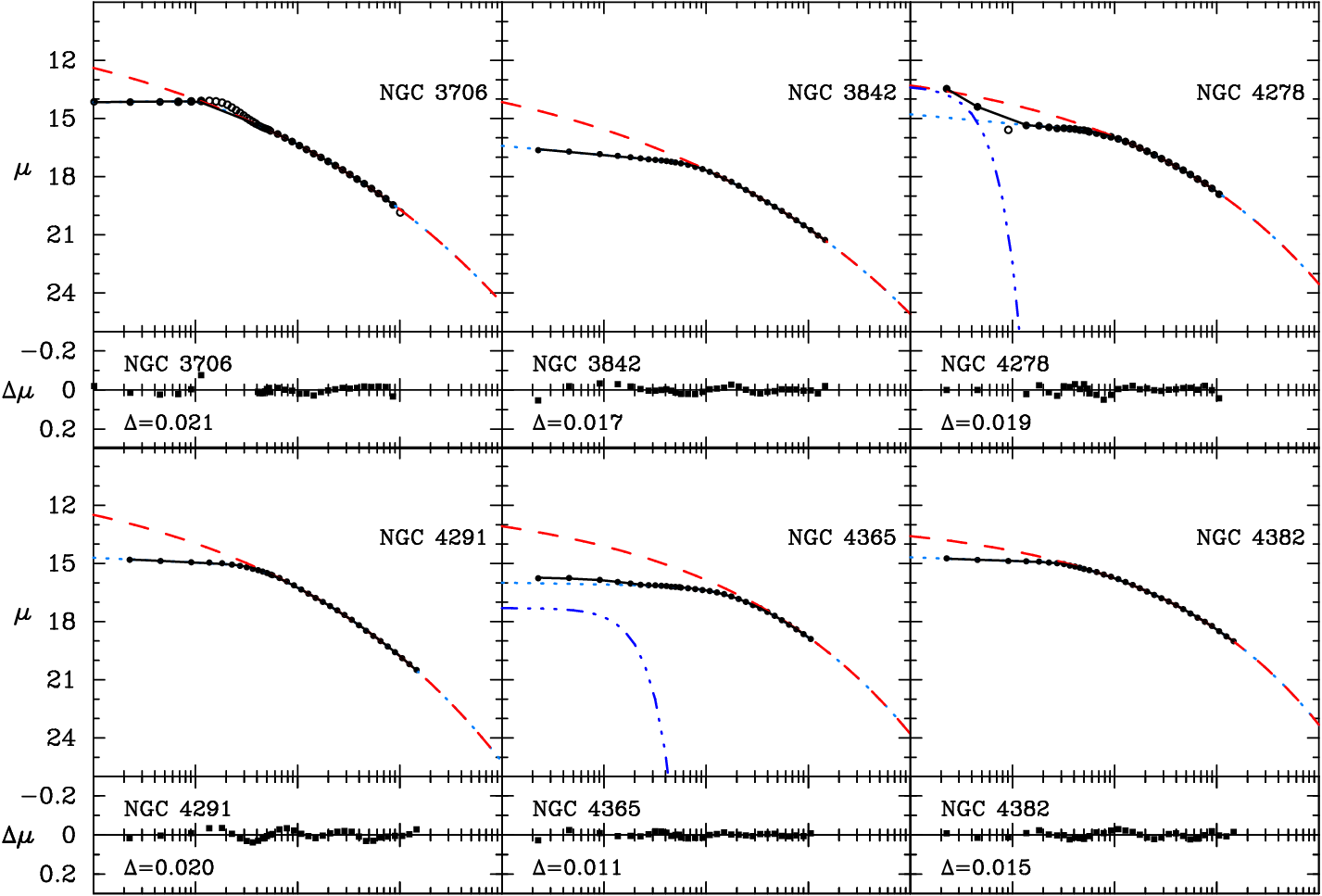}
\includegraphics[angle=270, scale=.68]{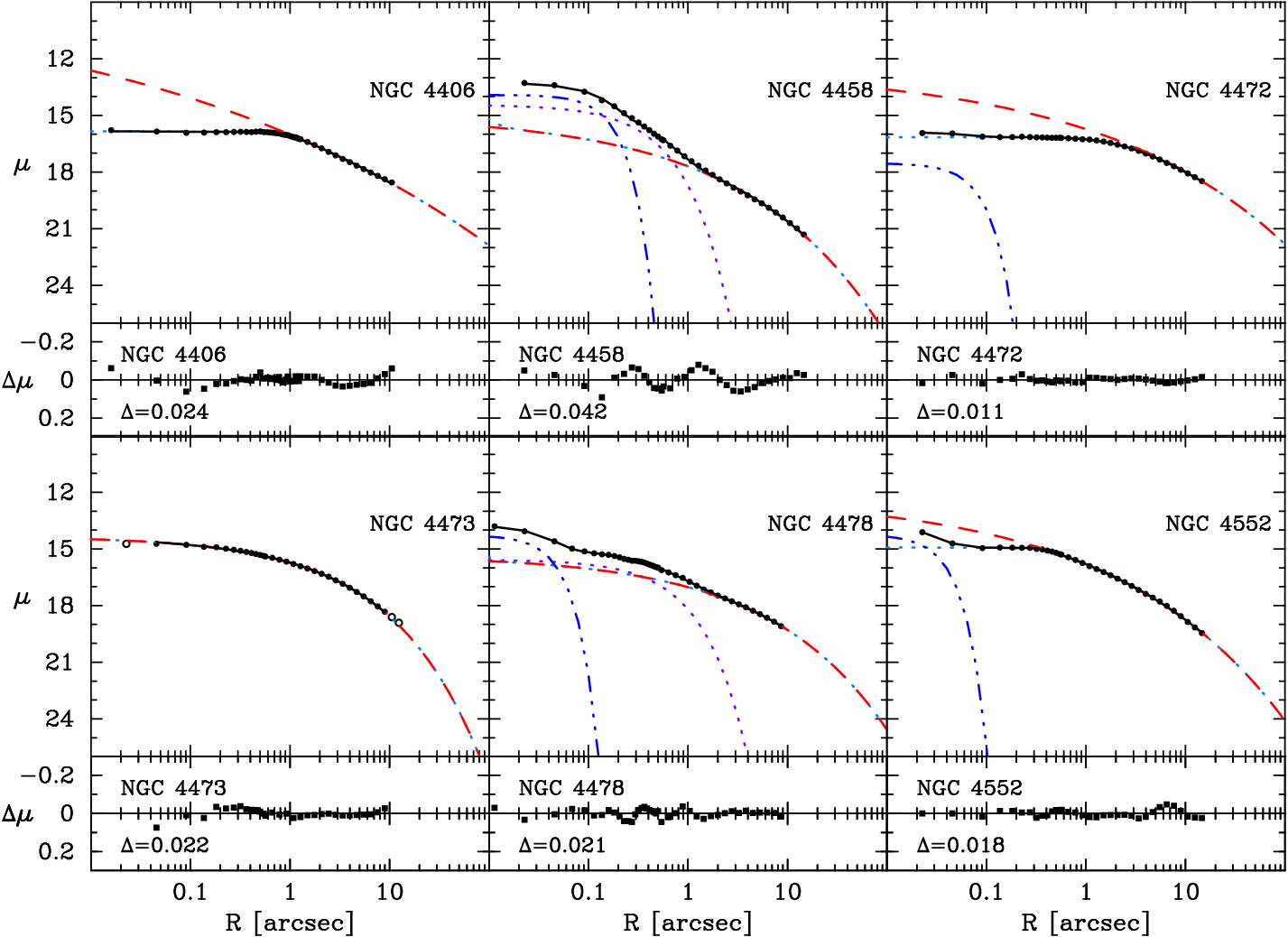}
 \caption{continued, see Appendix B text for further details.}
\end{figure*}
\end{center}

\setcounter{figure}{21}
\begin{center}
\begin{figure*}
\includegraphics[angle=270, scale=.68]{grp5.ps}
\includegraphics[angle=270, scale=.68]{grp6.ps}
\caption{continued, see Appendix B text for further details.}
\end{figure*}
\end{center}

\begin{center}
\begin{figure*}
\includegraphics[angle=270,scale=.68]{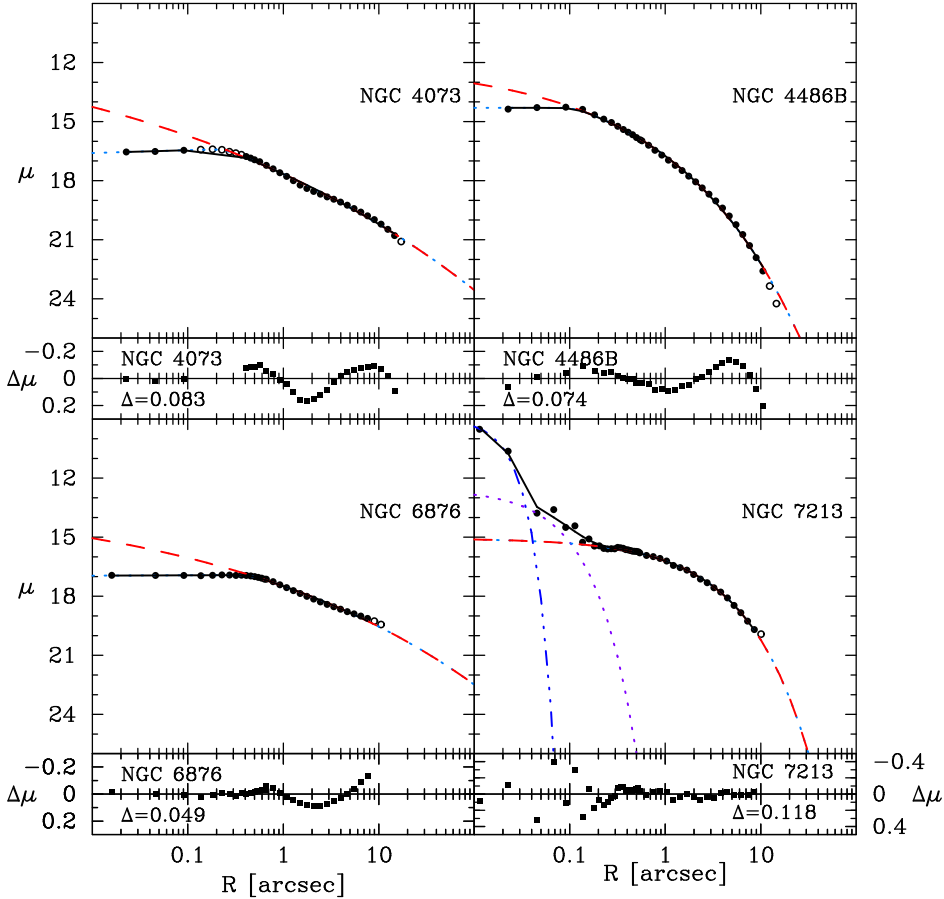}
\caption{Similar to Fig.~\ref{Fig17}, but for galaxies with various morphological peculiarities. See Appendix B text for further details.}
\label{Fig18}
\end{figure*}
\end{center}~~~~

\section{Appendix B}

\subsection{Notes on selected individual galaxies}

\noindent
{\bf NGC 584:} This galaxy has a depleted core with an inner negative
logarithmic slope $\gamma = 0.46$. Thus, it has $0.3 < \gamma < 0.5$ as
opposed to the core ($\gamma < 0.3$) / power-law ($\gamma > 0.5$) dichotomy
(see also Glass et al.\ 2011).

\noindent
{\bf NGC 1374:} This galaxy hosts an elongated inner disk (van der Marel \&
Franx 1993), which is well fit with a S\'ersic model plus an inner exponential
function for representing the disk.

\noindent
{\bf NGC 3607:} There is vivid evidence for the presence of an inner dust ring
(Lauer et al.\ 2005).

\noindent
{\bf NGC 3640:} The galaxy is known for its morphological peculiarity
including several sharp and diffuse features (e.g.\ shells, ripples), which
probably signal an ongoing or recent merger (Michard \& Prugniel 2004), which
makes the underlying host galaxy light difficult to model since our model is
not designed to accommodate such peculiarities. An ongoing or recent merger
could be the reason for having an unusually small core for a galaxy of this
luminosity.

\noindent
{\bf NGC 3706:} This galaxy exhibits an obvious inner stellar ring, shifting
the peak of the surface brightness to $R\approx 0\arcsec.14$ from the
photometric center (Lauer et al\ 2005). The light profile is better modelled
with careful omission of the additional ring of starlight from 0\arcsec.11 -
0\arcsec.4 (see also Capetti \& Balmaverde 2005).

\noindent
{\bf NGC 4073:} A cD galaxy with a possible history of cannibalism and having
similar distinct features as NGC 3706. Lauer et al.\ (2005) identified an
asymmetric central ring of stars noticeable on the brightness profile of the
galaxy. Data points from $0\arcsec.1 $ - $0\arcsec.4$ are excluded from the
fit.

\noindent
{\bf NGC 4365:} This object is a giant elliptical in the Virgo cluster with a
kinematically distinct core (Forbes et al.\ 1994), that has a slightly
elongated inner nuclear excess (Carollo et al.\ 1997; C\^ot\'e et al.\
2006). The surface brightness profile is well fitted with the core-S\'ersic
model plus a Gaussian function for the inner nucleus.

\noindent
{\bf NGC 4458:} The isophotal contour analysis presented in  Trujillo et al's.\ (2004) revealed
a prominent nuclear disk, see also Morelli et al.\ (2004). Analysis of the
brightness profile indicates the presence of an extended point source
(Ferrarese et al.\ 2006) and an inner disk.  A good match to the
\emph{HST}-observed profile is obtained fitting a S\'ersic model plus an inner
exponential ($n=1)$ disk component and a Gaussian ($n=0.5$) for the point
source.

\noindent
{\bf NGC 4478:} Like NGC 4458, there evidence for the presence of a nuclear
disk in this galaxy (e.g.\ Trujillo et al.\ 2004; Morelli et al.\ 2004). The
only noticeable difference with the light profile of NGC 4458 is that this
galaxy has a relatively elongated disk, and a compact point source (Carollo et
al.\ 1997; Ferrarese et al.\ 2006).

\noindent
{\bf NGC 4486B:} Based on the deconvolved \emph{HST}/WFPC2 \emph{I}- and
\emph{V}-band images, Lauer et al.\ (1996) showed the existence of a double
optical nucleus in this galaxy. However, the double nuclei are not obvious
from most of the archival optical \emph{HST} images, as noted by Ferrarese et
al.\ (2006), and it may be an artifact of the deconvolution
routine. Nonetheless, the plateau in the inner light profile is not due to a
relative deficit of stars. Tidal truncation from the interaction with the
close companion M87 is apparent from the fit to the surface brightness
profile, particularly in the outer part of the profile ($R >$ 7$\arcsec
\approx 0.7$ kpc).

\noindent
{\bf NGC 4552:} This galaxy hosts a compact point source which is detectable
in the optical image (Renzini et al.\ 1995; Carollo et al.\ 1997; Cappellari et al.\ 1999; Lauer et
al.\ 2005).

\noindent
{\bf NGC 5419:} There seem to be two compact nuclear point sources, the
brighter one at the photometric center, visible in the optical images (e.g.\
Capetti \& Balmaverde 2005; Lauer et al.\ 2005). A ``dip", in the region
0\arcsec.1 to 0\arcsec.2, is detected in the light profile of the
galaxy. Thus, two data points from $0\arcsec.1 < R < 0\arcsec.2$ are excluded
from the fit.

\noindent
{\bf NGC 6876:} The dominant elliptical in the Pavo group shows past or
ongoing interaction with the large spiral NGC 6872 (Machacek et al.\
2005). Furthermore, the archival \emph{I}- and \emph{V}-band images of this
galaxy indicate the presence of a double optical nucleus (Fig.~\ref{Fig3}),
possibly the semi-digested nuclei of lesser galaxies or
 the ends of an inclined ring (Lauer et al.\ 2002). 
Like NGC~4486B, a core-S\'ersic model can fit the surface
brightness profile (see Fig.~\ref{Fig18}).  Although, as the nuclei appear
only $0\arcsec.16$, and equidistant, from the center, they may not explain the
core's structure and break radius of 0.45$\arcsec$.  

\noindent
{\bf NGC 7213:} This galaxy is the only spiral (Sa) in our sample. It hosts a
bright Seyfert nucleus (V\'eron-Cetty \& V\'eron 1988) and an inner
disk. Hameed et al.\ (2001), see also Grosb{\o}l et al.\ (2004), argue, based
on the HI (neutral hydrogen) map, that this a highly disturbed system that may
have experienced a past merger. In addition, using the broadband \emph{HST}
images, Deo, Crenshaw \& Kraemer (2006) noted nuclear dust features in this
galaxy.

\label{lastpage}

\begin{references}
\reference{Allen}Allen, P.D., Driver, S.P., Graham, A.W., Cameron, E., Liske, J., De Propris, R., 2006, MNRAS, 371, 2
\reference{Andre} Andredakis Y. C., Peletier, R. F., \& Balcells, M. 1995, MNRAS, 275, 874
\reference{bal1} Balcells, M., Graham, A. W., Dom\'ingez-Palmero L., \& Peletier, R. F.\ 2003,
ApJ, 582, L79
\reference{BaC06}Balmaverde, B., Capetti A., 2006, A\&A, 447, 97
\reference{begl} Begelman, M. C., Blandford, R. D., \& Rees, M. J.\
1980, Nature, 287, 307
\reference{Bell4} Bell, E.F., Wolf C., Meisenheimer K., et al.\ 2004, ApJ, 608, 752 
\reference{bekki} Bekki, K. \& Graham, A. W. 2010, ApJ, 714, L313
\reference{Ber02} Bernardi, M., et al.\ 2002, AJ, 123, 2990
\reference{Bin84}Binggeli, B., Sandage, A., Tarenghi, M.\ 1984, AJ, 89, 64
\reference{BaM82}Binney, J., Mamon, G.A.\ 1982, MNRAS, 200, 361
\reference{birre} Biretta, J., et al.\ 2001, WFPC2 Instrument Handbook, Version 6.0 (Baltimore: STScI)
\reference{black} Blakeslee, J. P., Lucey, J. R., Tonry, J. L., Hudson, M. J., Narayanan, V. K.,
Harris, B. J., 2002, MNRAS, 330, 443
\reference{byun} Byun, Y.-I., et al.\ 1996, AJ, 111, 1889
\reference{caon} Caon, N., Capaccioli, M., D'Onofrio, M.\ 1993, MNRAS, 265, 1013
\reference{capn} Capetti, A., Balmaverde, B., 2005, A\&A, 440, 73
\reference{cappe} Cappellari, M., Renzini, A., Greggio, L., di Serego Alighieri, S., Buson, L. M.,
Burstein, D., Bertola, F., 1999, ApJ, 519, 117
\reference{carol} Carollo, C. M., Franx, M., Illingworth, G. D., \& Forbes, D. A.\ 1997, ApJ,
481, 710
\reference{con98} Condon, J.J., et al.\ 1998, AJ, 115, 1693
\reference{con02} Condon, J.J., et al.\ 2002, yCat, 8065
 \reference{cote4} C\^ot\'e, P., et al., 2006, ApJS, 165, 57
\reference{cote} C\^ot\'e, P., et al.\ 2007, ApJ, 671, 1456
\reference{cox} Cox, T. J., Jonsson, P., Primack, J. R., \& Somerville, R. S. 2006, MNRAS,
373, 1013
\reference{cran} Crane, P., et al.\ 1993, AJ, 106, 1371
\reference{dhar} Dhar, B. K., \& Williams, L. L. R. 2010, MNRAS, 405, 340
\reference{dhar1} Dhar, B. K., \& Williams, L. L. R., arXiv:1112.3120 
\reference{Davie} Davies, R.L., Efstathiou, G., Fall, S.M., Illing- worth, G., Schechter, P.L. 1983, ApJ, 266, 41
\reference{deo} Deo, R. P., Crenshaw, D. M., \& Kraemer S. B. 2006, AJ, 132, 321
\reference{derr} de Ruiter, H. R., Parma, P., Capetti, A., Fanti, R., Morganti, R., Santantonio, L.,
2005, A\&A, 439, 487
\reference{dev48} de Vaucouleurs, G., 1948, Ann. d'Astrophys., 11, 247
\reference{deV91} de Vaucouleurs, G., de Vaucouleurs, A., Corwin, H.G., Jr.,
et al.\ 1991, Third Reference Catalogue of Bright Galaxies, Springer-Verlag Berlin Heidelberg New York
\reference{diehl} Diehl, S., \& Statler, T. S., 2008, ApJ, 680, 897
\reference{djorg} Djorgovski, S., Davis, M., 1987, ApJ, 313, 59
\reference{DCC94}D'Onofrio, M., Capaccioli, M., Caon, N.\ 1994, MNRAS, 271, 523
\reference{faber} Faber, S. M., et al.\ 1997, AJ, 114, 1771
\reference{ferr06} Ferrarese, L., et al.\ 2006, ApJS, 164, 334
\reference{ferr00} Ferrarese, L., Merritt, D., 2000, ApJ, 539, L9
\reference{ferra1} Ferrarese, L., van den Bosch, F. C., Ford, H. C., Jaffe, W., \& O'Connell, R. W.\
1994, AJ, 108, 1598
\reference{FDCNP}Ferrari, F., Dottori, H., Caon, N., Nobrega, A., Pavani, D.B.\ 2004, MNRAS, 347, 824
\reference{forb1} Forbes, D. A., 1994, AJ, 107, 2017
\reference{fuk} Fukugita, M., Shimasaku K., \& Ichikawa, T. 1995, PASP, 107, 945
\reference{Gav05}Gavazzi, G., Donati, A., Cucciati, O., Sabatini, S., Boselli, A., Davies, J., Zibetti, S.\ 2005, A\&A, 430, 411
\reference{geb} Gebhardt, K., et al.\ 1996, AJ, 112, 105 
\reference{geb200} Gebhardt, K., Bender R., Bower G., et al.\ 2000, ApJ, 539, L13 
\reference{grawdd} Gebhardt, K., et al., 2007, ApJ, 671, 1321
\reference{gra6} Glass, L., et al., 2011, ApJ, 726, 31
\reference{Goer} Goerdt, T., Moore, B., Read, J. I., Stadel, J.\ 2010, ApJ, 725, 1707
 \reference{gra04} Graham, A.W., 2004, ApJ, 613, L33
 \reference{gra07} Graham, A.W., 2007, MNRAS, 379, 711
 \reference{gra8a} Graham, A.W., 2008a, ApJ, 680, 143
 \reference{gra8b} Graham, A.W., 2008b, Publ.\ Astron.\ Soc.\ Aust., 25, 167
 \reference{gr11a} Graham, A.W., 2012a, in ``Planets, Stars and Stellar Systems'', Springer Publishing (arXiv:1108.0997)
 \reference{gr11b} Graham, A. W., 2012b, ApJ, 746, 113
 \reference{gra05} Graham, A.W., Driver, S.P.\ 2005, Publ. Astron. Soc. Australia, 22, 118
\reference{gra03} Graham, A.W., Erwin P., Trujillo I., Asensio Ramos A.\ 2003, AJ, 125, 2951
 \reference{GaG03} Graham, A.W., Guzm\'an R.\ 2003, AJ, 125, 2936
 \reference{gra96} Graham, A.W., Lauer T.R., Colless M.M., Postman M.\ 1996, ApJ, 465, 534
 \reference{Graet} Graham, A.W., Onken C.A., Athanassoula E., Combes F.\ 2011, MNRAS, 412, 2211
\reference{GaW08} Graham, A.W., Worley C.C., 2008, MNRAS, 388, 1708 
 \reference{grill} Grillmair, C. J., et al.\ 1994, AJ, 108, 102
 \reference{grosb} Grosb{\o}l P., Patsis, P. A., \& Pompei, E., 2004, A\&A, 423, 849
\reference{Gual} Gualandris, A. \& Merritt, D. 2011, arXiv1107.4095
\reference{Gul09}G{\"u}ltekin, K., Richstone, D.~O., Gebhardt, K., et al.\ 2009, ApJ, 695, 1577
\reference{Gul11}G{\"u}ltekin, K., Richstone, D.~O., Gebhardt, K., et al.\ 2011, ApJ, 741, 38
\reference{haam} Hameed, S., Blank, D. L., Young, L. M., Devereux, N., 2001, ApJ, 546, L97
 \reference{hern} Hernquist, L.,\ 1990, ApJ, 356, 359
  \reference{hern2} Hernquist, L., 1993, ApJ, 409, 548
\reference{hopk} Hopkins, P. F., \& Hernquist, L. 2010, MNRAS, 407, 447
\reference{jaffe} Jaffe, W., Ford H. C., O'Connell, R. W., van den Bosch, F. C., \& Ferrarese, L.\ 1994, AJ, 108, 1567
\reference{kandrup} Kandrup, H. E., Sideris, I. V., Terzi\'c, B., Bohn, C. L., 2003, ApJ, 597, 111
\reference{Khoch}Khochfar, S., Burkert, A., 2003, ApJ, 597, 117 
\reference{Kin78}King, I.R.\ 1978, ApJ, 222, 1
\reference{KaM66}King, I.R., Minkowski, R.\ 1966, ApJ, 143, 1002
\reference{KaM72}King, I.R., Minkowski, R.\ 1972, IAU Symp., 44, 87
\reference{korm2} Kormendy, J., \& Bender, R. 2009, ApJ, 691, L142
\reference{korm3} Kormendy, J., Dressler, A., Byun, Y. I., Faber, S. M., Grillmair,
  C., Lauer, T. R., Richstone, D., Tremaine, S.\ 1994, in ESO/OHP Workshop on
  Dwarf Galaxies, ed. G. Meylan \& P. Prugniel (Garching: ESO), 147
\reference{Kormen199} Kormendy, J., 1999, in Merritt D. R., Valluri M., Sellwood J. A., eds, ASP
Conf. Ser. Vol. 182, Galaxy Dynamics -A Rutgers Symposium, Astron.
Soc. Pac., San Francisco, p. 124
\reference{Kor09}Kormendy, J., Fisher, D.B., Cornell, M.E., Bender, R.\ 2009, ApJS, 182, 216
\reference{Kulkarni} Kulkarni, G. \& Loeb A., arXiv1107.0517
\reference{lain} Laine, S., van der Marel, R. P., Lauer, T. R., Postman, M., ODea, C. P., \&
Owen, F. N. 2003, AJ, 125, 478
\reference{laer2} Lauer, T. R., et al.\ 1995, AJ, 110, 2622
\reference{laehh} Lauer, T. R., et al.\ 1996, ApJ, 471, L79
\reference{Lau02} Lauer, T.~R., Gebhardt, K., Richstone, D., et al.\ 2002, AJ, 124, 1975 
\reference{laer3} Lauer, T. R., et al.\ 2005, AJ, 129, 2138
\reference{lau9} Lauer, T. R., et al., 2007a, ApJ, 662, 808
\reference{lawq} Lauer, T. R., et al., 2007b, ApJ, 664, 226
\reference{laurk} Laurikainen, E., Salo, H., Buta, R., Knapen, J. H., \& Comer\'on, S. 2010, MNRAS, 405, 1089
\reference{lotz} Lotz, J. M., Miller, B. W., \& Ferguson, H. C. 2004, ApJ, 613, 262
\reference{lywq} Lyubenova M., Kuntschner, H., Silva, D. R., 2008, A\&A, 485, 425
\reference{Mach5} Machacek M., Nulsen, P. E.J., Stirbat, L., Jones, C., Forman, W.R.\ 2005, ApJ, 630, 280
\reference{makin} Makino, J., \& Ebisuzaki, T.\ 1996, ApJ, 465, 527
\reference{marc} Marconi, A., Hunt, L.K. 2003, ApJ, 589, L21
\reference{martiz} Martizzi, D., Teyssier, R. and Moore, B.\ 2012, MNRAS, 420, 2859
\reference{mikrr} Merritt, D., 2006, ApJ, 648, 976
\reference{merrit1} Merritt, D., \& Milosavljevi\'c, M.\ 2005, Living Rev. Relativ., 8, 8
\reference{mika} Michard, R., Prugniel, P., 2004, A\&A, 423, 833
\reference{mikall} Milosavljevi\'c, M., \& Merritt, D. 2001, ApJ, 563, 34
 \reference{mor} Morelli, L. et al., 2004, MNRAS, 354, 753
\reference{NJB06} Naab, T., Jesseit, R., Burkert, A., 2006, MNRAS, 372, 839
 \reference{Na)09} Naab, T., \& Ostriker, J. P. 2009, ApJ, 690, 1452
 \reference{Mak99} Nakano, T., \& Makino, J. 1999, ApJ, 510, 155
\reference{nieto} Nieto, J.-P., \& Bender, R. 1989, A\&A, 215, 266
\reference{nietos} Nieto, J.-L., Bender, R., Surma, P., 1991, A\&A, 244, L37
\reference{pat} Paturel, G., Petit, C., Prugniel, P., Theureau, G., Rousseau, J., Brouty, M.,
Dubois, P., Cambr\'esy, L., 2003, A\&A, 412, 45
\reference{pel10} Pellegrini, S.\ 2010, ApJ, 717, 640
\reference{ravi} Ravindranath, S., Ho, L. C., Peng, C. Y., Filippenko, A. V., \& Sargent, W. L. W.\ 2001. AJ, 122, 653
\reference{Renz5}Renzini, A., Greggio, L., di Serego Alighieri, S., et al.\ 1995, Nature, 378, 39
\reference{rest} Rest, A., van den Bosch, F. C., Jaffe, W., Tran, H., Tsvetanov, Z., Ford, H. C.,
Davies, J., \& Schafer, J.\ 2001, AJ, 121, 2431
\reference{riching} Richings, A. J., Uttley, P., \& Kr$\ddot{\textrm{o}}$ding, E., 2011, MNRAS, tmp, 759
\reference{sersic1} S\'ersic, J. L.\ 1963, Bolet\'in de la Asociaci\'on Argentina de Astronom\'ia, 6, 41
\reference{sersic2}S\'ersic, J. L.\ 1968, Atlas de Galaxias Australes (Cordoba: Observatorio
Astronomico)
\reference{SVP03}Soldatenkov, D.~A., Vikhlinin, A.~A., \& Pavlinsky, M.~N.\ 2003, Astronomy Letters, 29, 298
\reference{some} Sommerville, R. S., Primack, J. R., 1999, MNRAS, 310, 1087
\reference{sori} Soria, R., Graham, A.W., Fabbiano, G., Baldi, A., Elvis, M., Jerjen,
H., Pellegrini, S., Siemiginowska, A., 2006, ApJ, 640, 143
\reference{tor} Tonry, J. L., et al., 2001, ApJ, 546, 681
\reference{tooo} Toomre, A., \& Toomre, J. 1972, ApJ, 178, 623
\reference{trui} Trujillo, I., Erwin, P., Asensio Ramos, A., \& Graham, A. W.\ 2004, AJ, 127,
1917
\reference{van94} van den Bosch, F. C., Ferrarese, L., Jaffe, W., Ford, H. C., \& O'Connell, R. W.\
1994, AJ, 108, 1579
 \reference{van93} van der Marel, R. P., Franx, M., 1993, ApJ, 407, 525
 \reference{vero} V\'eron-Cetty, M.-P., \& V\'eron, P. 1988, A\&A, 204, 28
\reference{WaB92}White, R.L., Becker, R.H., 1992, ApJS, 79, 331
\reference{Worth}Worthey, G. 1994, ApJS, 95, 107
\reference{Xu011}Xu, C.K., Zhao, Y., Scoville, N., Capak, P., Drory, N., Gao, Y.\ 2011, ApJ, submitted (ArXiv:1109.3693)
\reference{young} Young, C. K., \& Currie, M. J. 1994, MNRAS, 268, L11
\reference{youngs} Younes, G., Porquet, D., Sabra, B., Grosso, N., Reeves, J. N., Allen, M. G.,
2010, A\&A, 517, 33
\reference{Yet78}Young, P.J., Westphal, J.A., Kristian, J., Wilson, C.P., Landauer, F.P.\ 1978, ApJ, 221, 721
\end{references}
\end{document}